\documentclass{aastex63}
\usepackage{float}
\makeatletter
\let\newfloat\newfloat@ltx
\makeatother
\usepackage{bm}
\usepackage{csquotes}
\usepackage{physics}
\usepackage{verbatim}
\usepackage{amsmath}
\usepackage{mathrsfs}
\usepackage{algorithm}
\usepackage{algorithmic}
\usepackage{xcolor}
\accepted{December 22, 2023}
\submitjournal{ApJ}

\shorttitle{Coronagraph modes post-processing}
\shortauthors{Xin et al.}
\graphicspath{{./}{figures/}}

\begin{document}

\title{Coronagraphic Data Post-processing Using Projections on Instrumental Modes}

\correspondingauthor{Yinzi Xin}
\email{yxin@caltech.edu}

\author[0000-0002-6171-9081]{Yinzi Xin}
\affiliation{Department of Aeronautics and Astronautics, Massachusetts Institute of Technology, Cambridge, MA 02139, USA}
\affiliation{Space Telescope Science Institute, 3700 San Martin Drive, Baltimore, MD, 21218, USA}
\affiliation{Department of Physics, California Institute of Technology, Pasadena, CA, 91125, USA}

\author[0000-0003-3818-408X]{Laurent Pueyo}
\affiliation{Space Telescope Science Institute, 3700 San Martin Drive, Baltimore, MD, 21218, USA}

\author[0000-0002-2215-9413]{Romain Laugier}
\affiliation{Université Côte d'Azur, 28 avenue Valrose, Nice, 06 06100, France}

\author[0000-0001-6387-9444]{Leonid Pogorelyuk}
\affiliation{Department of Aeronautics and Astronautics, Massachusetts Institute of Technology, Cambridge, MA 02139, USA}

\author[0000-0002-0813-4308]{Ewan S. Douglas}
\affiliation{University of Arizona, Steward Observatory, 933 N. Cherry Ave, Tucson, AZ 85721, USA}

\author[0000-0003-2595-9114]{Benjamin J. S. Pope}
\affiliation{School of Mathematics and Physics, The University of Queensland, St Lucia, QLD 4072, Australia}
\affiliation{Centre for Astrophysics, University of Southern Queensland, West Street, Toowoomba, QLD 4350, Australia}

\author[0000-0002-7791-5124]{Kerri L. Cahoy}
\affiliation{Department of Aeronautics and Astronautics, Massachusetts Institute of Technology, Cambridge, MA 02139, USA}



\begin{abstract}

Directly observing exoplanets with coronagraphs is impeded by the presence of speckles from aberrations in the optical path, which can be mitigated in hardware with wavefront control as well as in post-processing. This work explores using an instrument model in post-processing to separate astrophysical signals from residual aberrations in coronagraphic data. The effect of wavefront error (WFE) on the coronagraphic intensity consists of a linear contribution and a quadratic contribution. When either of the terms is much larger than the other, the instrument response can be approximated by a transfer matrix mapping WFE to detector plane intensity. From this transfer matrix, a useful projection onto instrumental modes that removes the dominant error modes can be derived. We apply this projection to synthetically generated Roman Space Telescope hybrid Lyot coronagraph (HLC) data to extract “robust observables," which can be used instead of raw data for applications such as detection testing. The projection improves planet flux ratio detection limits by about $28 \%$ in the linear regime and by over a factor of 2 in the quadratic regime, illustrating that robust observables can increase sensitivity to astrophysical signals and improve the scientific yield from coronagraphic data. While this approach does not require additional information such as observations of reference stars or modulations of a deformable mirror, it can and should be combined with these other techniques, acting as a model-informed prior in an overall post-processing strategy.

\end{abstract}



\section{Introduction} \label{sec:intro}

Specialist high-contrast techniques are required to directly observe faint astrophysical objects near brighter objects, such as exoplanets, brown dwarfs, or circumstellar disks orbiting much brighter central stars. High contrast observations are essential for answering scientific questions involving binary and planetary system population statistics, planet and disk formation and evolution, planetary atmospheres, and planet habitability and the search for biosignatures \citep{traub_oppenheimer}. Measuring these exoplanet signals is difficult because they often lie at small angular separations from their host star and can be many orders of magnitude fainter. One major obstacle for high contrast observations is photon noise from the light of the central star. Practical matters such as detector saturation aside, if the star is orders of magnitude brighter than its companion, the photon noise associated with the outer lobes of star's point-spread-function (PSF) can overwhelm any signal from the companion, even if the on-axis star's signal is perfectly known. As a result, instruments to directly suppress starlight, such as coronagraphs and nullers, are important in increasing the photon signal-to-noise (SNR) of faint companions. 

Another important source of noise is wavefront error (WFE), which distorts the signal of the on-axis point source. Sources of wavefront error include atmospheric turbulence, imperfections in the optics, or thermo-mechanical changes in the telescope or instrument. At an instant in time, a perturbation to the wavefront scatters energy from the core of the PSF into speckles throughout the image that can resemble off-axis sources. If the magnitude of the WFE electric field is smaller than that of the underlying electric field from the PSF, then the speckles are symmetric about zero in detector plane intensity and average out over time. When the wavefront error is the larger term, as in the case of uncorrected atmospheric turbulence, the speckles are predominantly positive, increasing rather than decreasing the intensity over most of the focal plane, and averaging out to a halo that can obscure off-axis signals. Scattered starlight at larger spatial separations increases the photon noise at those locations in the detector plane, which can also dominate over signals from faint companions.

The goal of high contrast instruments is to separate the signal of the on-axis star from off-axis sources. Coronagraphs are passive optical elements that spatially filter the light to suppress the signal of an on-axis star, reducing its associated photon noise while letting through off-axis signals \citep{Guyon_2006}. Adaptive optics \citep[AO;][]{tyson} and focal plane wavefront control \citep{groff_2015} actively correct for wavefront error to reduce their impact. However, even with suppression from coronagraphs or nullers, the sensitivity to faint astrophysical signals is still limited by residual starlight and its associated photon noise.

Post-processing techniques can use additional available information to further mitigate the effects of WFE and increase sensitivity to real astrophysical signals (see \citet{cantalloube_2022} for a discussion of the state-of-the-art of high contrast post-processing in the context of a direct imaging data challenge). For example, angular differential imaging (ADI) exploits observations at different roll angles, taking advantage of azimuthal averaging of the wavefront error \citep{Marois_2006,flasseur}. Other methods rely on performing principal component analysis (PCA) on reference observations of a calibration star similar to the host star, but without astrophysical companions, to calibrate out residual static or quasi-static starlight \citep{lafreniere_loci, soummer_klip, pueyo_klip}. Additional sources of information on residual WFE include telemetry from wavefront sensing and control (WFSC) systems such as wavefront sensor residuals \citep{Vogt_2011} or focal plane electric field estimates \citep{Pogorelyuk_2019}, data from a self-coherent camera \citep{Baudoz_2006}, and data at different wavelengths as exploited in spectral deconvolution \citep{Sparks_2002}. 

This work shows that the modeled or measured instrument sensitivity to wavefront error can be included as an additional source of information in the post-processing of coronagraphic data, information that, in theory, can be combined with the other techniques discussed. This work examines an approach that uses this physical optics model to construct a projection removing the dominant error modes in the appropriate wavefront error regime, and finds that this can improve sensitivity to faint companions by up to and over a factor of 2.

\section{Coronagraphic Signals}

\subsection{Data Formation} \label{sec:data_formation}
The model used in this work assumes the light through the instrument is monochromatic. With a discrete representation of the optical planes of an instrument, a coronagraph can be modeled as a linear operator $\mathbf{C}$, a constant 2D matrix transforming the electric field vector at the pupil plane, $\bm{E_{s}}$, into the electric field vector at the detector plane, $\bm{E_\textbf{det}}$.  If $\bm{E_{s0}}$ is the electric field vector of the central source (star) at the pupil plane in the absence of aberrations, and $\bm{\mathit{\Delta} E_s}$ a vector of small perturbations to that electric field, representing wavefront aberrations (which can be variable in time), then the electric field vector at the detector plane, assuming that the star is the only source of light, is

\begin{equation}
\bm{E_\textbf{det}} = \mathbf{C} \bm{E_{s}} = \mathbf{C} \bm{E_{s0}} + \mathbf{C} \bm{\mathit{\Delta} E_{s}}(t).
\end{equation}

The intensity measured is the element-wise norm squared of the detector plane electric field (here, $\overline{x}$ indicates the element-wise complex conjugate of $x$ and $\circ$ indicates the element-wise product):

\begin{equation} \label{eq:Is}
\begin{split}
\bm{I_s} & =  |\bm{E_{\textbf{det}}}|^2 \\
& =  |\mathbf{C} \bm{E_{s0}}|^2 + 2 \Re{\overline{(\mathbf{C} \bm{E_{s0}})} \circ \mathbf{C} \bm{\mathit{\Delta} E_s}(t)} + |\mathbf{C} \bm{\mathit{\Delta} E_s}(t)|^2.
\end{split}
\end{equation}

The vector of the pupil plane electric field of a binary companion is given by

\begin{equation}
\bm{E_p}  = \sqrt{c}\bm{E_{s0}}e^{-i\bm{k} \cdot \bm{x}}+ \sqrt{c}\bm{\mathit{\Delta} E_s}(t)e^{-i\bm{k} \cdot \bm{x}} = \bm{E_{p0}} + \bm{\mathit{\Delta} E_p}(t),
\end{equation}

where $c$ is the flux ratio between the planet and the star, $\bm{k}$ is the pupil plane wave vector indicating the companion's location, and $\bm{x}$ is the pupil plane coordinate vector. Namely, the planet's pupil-plane electric field is the star's electric field, but tilted and scaled by the square root of the flux ratio. The detector plane intensity for the planet can be expressed as

\begin{equation} \label{eq:Ip}
\bm{I_p} = |\mathbf{C} \bm{E_{p0}}|^2+2 \Re{ \overline{(\mathbf{C} \bm{E_{p0}})} \circ \mathbf{C} \bm{\mathit{\Delta} E_p}(t)} + |\mathbf{C} \bm{\mathit{\Delta} E_p}(t)|^2.
\end{equation}

The total intensity on the detector plane from the star and the planet is the sum of Eqs. \ref{eq:Is} and \ref{eq:Ip}. However, we can make two simplifying assumptions. The first assumption is that the flux of the planet is small relative to the flux of the star, such that $c \ll 1$. The second assumption is that the magnitude of the wavefront error is small relative to the total magnitude of the electric field, namely $\bm{\mathit{\Delta} E_s}(t) \ll \bm{E_{s0}}$, which implies $\bm{\mathit{\Delta} E_p}(t) \ll \bm{E_{p0}}$. This is true if we are both in the ``small phase regime" (when there is much less than one wave of wavefront error) and the fractional amplitude error is much less than one. These assumptions imply that the last two terms of Eq. \ref{eq:Ip} are small relative to the other terms, so we can approximate the total intensity as

\begin{equation} \label{eq:Itot}
\bm{I_\textbf{tot}}  \approx  |\mathbf{C} \bm{E_{s0}}|^2 + 2 \Re{\overline{(\mathbf{C} \bm{E_{s0}})} \circ \mathbf{C} \bm{\mathit{\Delta} E_s}(t)} + |\mathbf{C} \bm{\mathit{\Delta} E_s}(t)|^2+ |\mathbf{C} \bm{E_{p0}}|^2.
\end{equation}

The first term $|\mathbf{C} \bm{E_{s0}}|^2$ is the residual starlight not blocked by the coronagraph in the case of no aberrations. The second term $2 \Re{\overline{(\mathbf{C} \bm{E_{s0}})} \circ \mathbf{C} \bm{\mathit{\Delta} E_s}}$ is linear in the wavefront aberration and corresponds to the interference of the aberration, propagated to the focal plane, with the underlying residual starlight from the coronagraph, analogous to speckle pinning \citep{Bloemhof_2002, perrin_2003}. The third term $|\mathbf{C} \bm{\mathit{\Delta} E_s}|^2$ is the quadratic term, corresponding to the norm squared of the wavefront error propagated to the focal plane. The last term $|\mathbf{C} \bm{E_{p0}}|^2$ is the nominal off-axis signal of interest.

Whether the effect of wavefront errors at some location in the detector plane are dominated by the linear term or the quadratic term depends on the attenuation of starlight by the coronagraph and the level of the propagated wavefront error at that location. If the propagated wavefront aberrations are smaller in complex amplitude than the residual starlight after the coronagraph with no aberrations, the linear term is dominant. When a coronagraph is not used, this corresponds to the speckle pinning regime, in which the aberrations primarily interfere with the wings of the telescope's PSF \citep{Bloemhof_2002}. The same phenomenon occurs with a coronagraph; however, as the amplitude of the PSF wings are reduced by the coronagraph, the range of WFE over which this occurs is much more limited. Otherwise, if the propagated wavefront aberrations have relatively larger magnitudes, the quadratic term is dominant. For a given location in the focal plane, the local point of transition between the linear and quadratic regimes occurs when $|2 \Re{\overline{(\mathbf{C} \bm{E_{s0}})} \circ \mathbf{C} \bm{\mathit{\Delta} E_s}}| = |\mathbf{C} \bm{\mathit{\Delta} E_s}|^2$, or roughly when $|2\mathbf{C} \bm{E_{s0}}| = |\mathbf{C} \bm{\mathit{\Delta} E_s}|$.

This point of transition is different for each pixel, and also depends on the coronagraph design as well as the ``nominal" wavefront (whether it is flat, as is typical for ground-based coronagraphs, or the wavefront corresponding to a dark hole, as is planned for space-based coronagraphs). For this work, we use as an example the Hybrid Lyot Coronagraph (HLC) of the Coronagraph Instrument of the Roman Space Telescope. With the dark hole presented in Section \ref{sec:hlc_model}, which has an average raw contrast (residual stellar intensity divided by unocculted peak intensity) of $5.6 \times 10^{-9}$, the point at which $|\mathbf{C} \bm{\mathit{\Delta} E_s}| > |2\mathbf{C} \bm{E_{s0}}|$ for $50\%$ of the pixels in the dark hole region occurs at roughly 0.1 waves root-mean-square (RMS) of phase error, on average. This means that wavefront errors less than 0.1 waves RMS will primarily be in the linear regime, while wavefront errors larger than 0.1 waves RMS will primarily be in the quadratic regime, although this is somewhat dependent on the form of the wavefront's spatial power spectral density (PSD) that we use in Section \ref{sec:syndata_gen}.

In this work, robust observables are only formulated for WFE that is predominantly linear or predominantly quadratic throughout the entire focal plane. However, it may be possible to obtain robust observables for when both terms have comparable contributions, a topic that is left for future work.

\subsection{Linear Regime}
From Equation \ref{eq:Itot}, if we then assume that the linear error term is dominant, then we can drop the quadratic contribution such that the detector plane intensity is approximately

\begin{equation} \label{eq:I_lin}
\bm{I_{\textbf{tot},l}}  \approx  |\mathbf{C} \bm{E_{s0}}|^2  + 2 \Re{\overline{(\mathbf{C} \bm{E_{s0}})} \circ \mathbf{C} \bm{\mathit{\Delta} E_s}(t)} + |\mathbf{C} \bm{E_{p0}}|^2.
\end{equation}

The contribution of the wavefront error to the intensity can be expressed as a linear transformation $\mathbf{A_l}$ acting on the wavefront error:

\begin{equation}
\bm{I_{\textbf{tot},l}}  \approx  |\mathbf{C} \bm{E_{s0}}|^2  + \mathbf{A_l} \bm{\mathit{\Delta} E_s}(t) + |\mathbf{C} \bm{E_{p0}}|^2.
\end{equation}

The transfer matrix $\mathbf{A_l}$ can be calculated semi-analytically from the coronagraph operator and the unaberrated electric field, as derived from Equation \ref{eq:I_lin}:

\begin{equation} \label{eq:A_l}
    \mathrm{A_l}_{kj} = \pdv{I_k}{\mathit{\Delta} E_{s_j}} = 2 \Re{(\sum_i \mathrm{C}_{ki} E_{s0_{i}})^* \mathrm{C}_{kj}}.
\end{equation}

The indices $i$ and $j$ label the input basis vectors used to represent the wavefront error, and the index $k$ labels the detector pixel.

It is desirable to reduce the term dependent on WFE, $\mathbf{A_l} \bm{\mathit{\Delta} E_s}(t)$, relative to the terms containing astrophysical signals of interest. This can be achieved by left-multiplying the measured intensities by a matrix $\mathbf{K_l}$, that projects out the dominant modes of $\mathbf{A_l}$. The following Section \ref{finding_k} describes the process of calculating $\mathbf{A_l}$ and finding from it an appropriate $\mathbf{K_l}$. The observables obtained using projection matrix $\mathbf{K_l}$ are given by

\begin{equation}
\bm{O_l} = \mathbf{K_l} \bm{I_{\textbf{tot},l}}.
\end{equation}

When the wavefront errors are in the linear regime, this projection is expected to suppress the contribution of wavefront errors to the measured data. As long as the measurements retain most of the astrophysical signal, then the projection will boost its SNR.

\subsection{Quadratic Regime}
In the quadratic-dominated regime, we can drop the linear contribution in Eq. \ref{eq:Itot}, such that the detector plane intensity is approximately

\begin{equation}
\bm{I_{\textbf{tot},q}}  \approx  |\mathbf{C} \bm{E_{s0}}|^2  + |\mathbf{C} \bm{\mathit{\Delta} E_s}(t)|^2+ |\mathbf{C} \bm{E_{p0}}|^2.
\end{equation}

In a discrete numerical model, the contribution of the quadratic term to each pixel labeled $k$ in the detector plane can be expressed as

\begin{equation} \label{eq:quad_tensor}
|\mathbf{C} \bm{\mathit{\Delta} E_s}|^2_k = (\sum_m \mathrm{C}_{km} \mathit{\Delta} E_{s_m})^*(\sum_n \mathrm{C}_{kn} \mathit{\Delta} E_{s_n}) = \sum_i \sum_j {\mathit{\Delta} E_{s_i}^*} \mathrm{M}_{kij}{\mathit{\Delta} E_{s_j}}.
\end{equation}

The indices $m$, $n$, $i$, and $j$ label the input basis vectors used to represent the wavefront error (expressed here in terms of perturbation to the complex electric field), and the index $k$ labels the detector pixel. The quantity $\mathbf{\hat{M}}$ with elements $\mathrm{M}_{kij}$ is a 3-tensor containing the second order partial derivative matrix (Hessian) of each pixel intensity with respect to the wavefront error, and relates each pairwise combination of pupil basis vectors to its effect on each detector plane pixel $k$. Each entry can be calculated semi-analytically from the coronagraph operator using the following formula derived from Equation \ref{eq:quad_tensor}:

\begin{equation} \label{eq:quad_tensor_elements}
    \mathrm{M}_{kij} = \pdv{I_k}{\mathit{\Delta} E_{s_i}^*}{\mathit{\Delta} E_{s_j}} = \mathrm{C}_{ki} \mathrm{C}_{kj}^*+\mathrm{C}_{kj} \mathrm{C}_{ki}^*.
\end{equation}

Assuming there are $N_{\text{pix}}$ pixels of interest on the detector, and $N$ basis vectors are used to represent the wavefront error, then, through a remapping, the 3-tensor $\mathbf{\hat{M}}$ of size ($N_{\text{pix}} \times N \times N$) can be expanded into a matrix acting on the space of all \textit{pairwise combinations} of pupil basis vectors. Since Hessians are symmetric because partial derivatives commute ($\mathrm{M}_{kij} = \mathrm{M}_{kji}$), the ordering of each pair of segments does not matter, and the derivatives corresponding to the same pair of original basis vectors can be consolidated into the same entry. This results in an input vector space of size ${N+1 \choose 2}$, or the number of pairwise combinations of pupil basis vectors. 

The 3-tensor $\mathbf{\hat{M}}$ can thus be represented as a ($N_{\text{pix}} \times {N+1 \choose 2}$) matrix $\mathbf{A_q}$ of second derivatives, acting on a vector $\bm{\beta}$ of perturbations defined for each pairwise combination ${\mathrm{\Delta} E_{s_i}}{\mathrm{\Delta} E_{s_j}}$. This results in the following expression for the quadratic term:

\begin{equation}
|\mathbf{C} \bm{\mathit{\Delta} E_s}|^2 = \mathbf{A_q} \bm{\beta}.
\end{equation}

The projection is similar to the linear case: the detector intensities can be left-multiplied by a matrix $\mathbf{K_q}$ that projects out the dominant quadratic error modes of $\mathbf{M}$. The observables with the appropriate projection $\mathbf{K_q}$ are given by

\begin{equation}
\bm{O_q} = \mathbf{K_q} \bm{I_{\textbf{tot},q}}.
\end{equation}

\section{Response Matrices and Robust Observables} \label{finding_k}

\subsection{Calculating the Response Matrix} \label{sec:response_matrices}

This section details the numerical calculation of instrument response matrices and the projection matrices. In this work, the response matrix is calculated with the wavefront aberrations represented in the Zernike basis. In this basis, $\mathit{\Delta} E_{Z_n}$ is the coefficient of the aberration induced by the $n^{th}$ Noll ordered Zernike polynomial \citep{Noll}, and $N$ is the total number of polynomials chosen to construct the response matrix:

\begin{eqnarray}
\bm{\mathit{\Delta} E_{s}} & = & \left(
\begin{array}{cc}
\mathit{\Delta} E_{Z_1}  \\ ... \\
\mathit{\Delta} E_{Z_N}
\end{array}\right).
\end{eqnarray}

We define $N_{\text{pix}}$ as the total number of detector pixels of the optical model and $N_{\text{basis}}$ as the number of Zernike modes to include. The coronagraph operator $\mathbf{C}$ is the $ N_{\text{pix}} \times N_{\text{basis}}$ matrix that, when applied to a vector of Zernike coefficients, gives the perturbation they induce in the focal plane electric field. This operator is typically either already part of the optical model, or obtainable by propagating Zernike modes through the optical model and using finite differences to populate its columns. Given the operator $\mathbf{C}$ and the initial unaberrated focal plane electric field, we can calculate both $\mathbf{A_l}$ and $\mathbf{A_q}$ using Equations \ref{eq:A_l} and \ref{eq:quad_tensor_elements}. Note that the term $(\sum_k \mathrm{C}_{kj} E_{s0_{j}})$ in Equation \ref{eq:A_l} is simply the initial unaberrated focal plane electric field at pixel $k$. For more complicated models without simple analytical solutions (such as those that include distortion), automatic differentiation, in which arbitrary exact derivatives can be computed without finite differences, may be useful \citep{pope2021}.

The linear transfer matrix poses no computational problems, as its size is $N_{\text{pix}} \times N_{\text{basis}}$. However, for the quadratic transfer matrix $\mathbf{A_q}$, the size of the input dimension quickly becomes computationally burdensome for high $N_{\text{basis}}$. For the example system shown in Section \ref{sec:example}, a $N_{\text{basis}}$ of 528 results in a $\mathbf{A_q}$ matrix of width ${N_{\text{basis}}+1 \choose 2} = 139,656 $ (the number of pairwise combinations of pupil basis vectors), and length 5,476 (the number of detector pixels of the model). This $\mathbf{A_q}$ matrix, when represented as (non-complex) doubles, is over 6GB in size. As explained in Section \ref{finding_k}, the calculation of the projection matrix involves a singular value decomposition (SVD) of the response matrix. Since calculating the SVD of a matrix of this size is too computationally expensive, we restrict our quadratic transfer matrix to only include the first $N_{\text{redu}}=100$ Zernikes, which results in an $\mathbf{A_q}$ with a width of only ${N_{\text{redu}}+1 \choose 2}  = 5050$. This model is valid only in a smaller area closer to the central star --- namely within $\sim 5\,\lambda/D$, where $\lambda$ is the wavelength and $D$ the telescope diameter. However, in Appendix \ref{app:a}, we explore using an approximation of the quadratic transfer matrix that can extend the area of applicability while circumventing impractical computational costs.

\subsection{Calculating the Projection Matrix}
Once a response matrix $\mathbf{A}$ is obtained, a singular value decomposition of $\mathbf{A} = \mathbf{U S V}^T$ is performed, revealing its singular modes and corresponding singular values. Then a choice of the number of modes to project out ($N_m$) is made. The remaining $N_\text{pix}-N_{m}$ modes are kept in the post-processing projection $\mathbf{K}$. Accordingly, $\mathbf{K}$ is the subset of $\mathbf{U}$ that contains the $m+1^{th}$ and higher left singular modes of $\mathbf{A}$. A pseudo-code summary of the process to find $\mathbf{K}$ is given in Algorithm \ref{alg:calc_K}. The optimal $N_m$ depends on the signal of interest. For the point-source companion signals explored in this work, $N_m$ is chosen as the cutoff that results in the best detection limit at the separation of interest.

\begin{algorithm}
\caption{Calculate projection matrix $\mathbf{K}$}
\textbf{Input:} Transfer matrix $\mathbf{A}$ \\
\textbf{Input:} Cutoff mode (number of modes to project out), $N_m$ \\
\textbf{Input:} Indices of detector plane pixels in region of interest, $\mathrm{idx}$ \\
\textbf{Output:} Projection matrix $\mathbf{K}$
\begin{algorithmic}

\label{alg:calc_K}
\STATE $U,S,V^T \leftarrow \mathrm{svd}(\mathbf{A})$
\STATE $\mathbf{K} \leftarrow \mathrm{transpose}(U(\mathrm{idx},N_m+1:\mathrm{end}))$

\end{algorithmic}
\end{algorithm}

If the linear and quadratic projections are used in the appropriate regimes to increase SNR, they could, for example, allow for a binary signal detection with a deeper flux ratio than using the raw intensity data. Detection tests can be performed on both projected and unprojected data to quantify this effect.

\section{Detection Testing} \label{sec:detection_tests}

Detections are typically claimed from a statistical hypothesis test \citep[see e.g.][]{Kasdin_2006, Jensen_Clem_2017,ceau}. A test statistic $T$ is calculated from the data and compared to a threshold $\xi$. A detection is claimed if $T \geq \xi$, and a lack of a detection is claimed otherwise. The fraction of real companions detected is the true positive rate ($\mathrm{TPR}$). A false positive occurs if there is no companion in the data, but the detection test incorrectly claims a detection. The rate at which this occurs is the false positive rate ($\mathrm{FPR}$).

As the detection threshold $\xi$ is decreased, detecting real companions becomes more likely, but false detections also become more likely \citep{Jensen_Clem_2017}. Varying the threshold and plotting the $\mathrm{TPR}$ as a function of the $\mathrm{FPR}$ results in a receiver operating characteristic (ROC) curve, an example of which is shown in Figure \ref{fig:hlc_hists}. ROC curves characterize the performance of a detection scheme and are used in the determination of flux ratio detection limits.

This work uses a simple Delta Reduced Chi-squared ($\mathit{\Delta} \chi^2_r$) statistic, or the difference in the reduced $\chi^2$ of the data assuming it contains only noise, and the reduced $\chi^2$ of the data assuming it contains noise and the companion signal. The formula for calculating this test statistic from the data is given by Equation \ref{eqn:delta_chi2r} (the bars indicate vector norm, the divisions are element-wise):

\begin{equation}
\label{eqn:delta_chi2r}
\mathit{\Delta} \chi^2_r = \frac{1}{\nu} \left(\left\lvert \frac{\mathbf{y}}{\bm{\sigma}} \right\rvert ^2 -\left\lvert \frac{\mathbf{y-x}}{\bm{\sigma}} \right\rvert ^2 \right).
\end{equation}

In this formula, $\mathbf{y}$ is the data, which is the synthetically generated realizations of $\bm{I_\textbf{tot}}$, with or without a planet. Meanwhile, $\mathbf{x}$ is the unaberrated model of the planet signal $\bm{I_{p0}} = |\mathbf{C} \bm{E_{p0}}|^2$ (assuming that it is known, such as through a maximum-likelihood-estimation). The estimated uncertainty of the data is denoted by $\bm{\sigma}$, and $\nu$ is the degrees of freedom (the number of data elements minus the number of free parameters; a binary system's three free parameters are the flux ratio, separation, and position angle). This use of this test statistic is motivated by an assumption that the noise is Gaussian and uncorrelated, under which this quantity is related to the relative log-probabilities of the data containing both the planet signal and noise, versus containing only noise. The noise being uncorrelated and Gaussian is generically not the case. However, the effects of the correlation and non-Gaussianity of the injected noise on the resulting test statistic distributions is properly simulated and captured by the Monte Carlo methods used in this work.

\section{Example: Nancy Grace Roman Space Telescope Hybrid Lyot Coronagraph}
\label{sec:example}

\begin{figure*}[!ht]
\begin{center}
	\includegraphics[scale = 2.0]{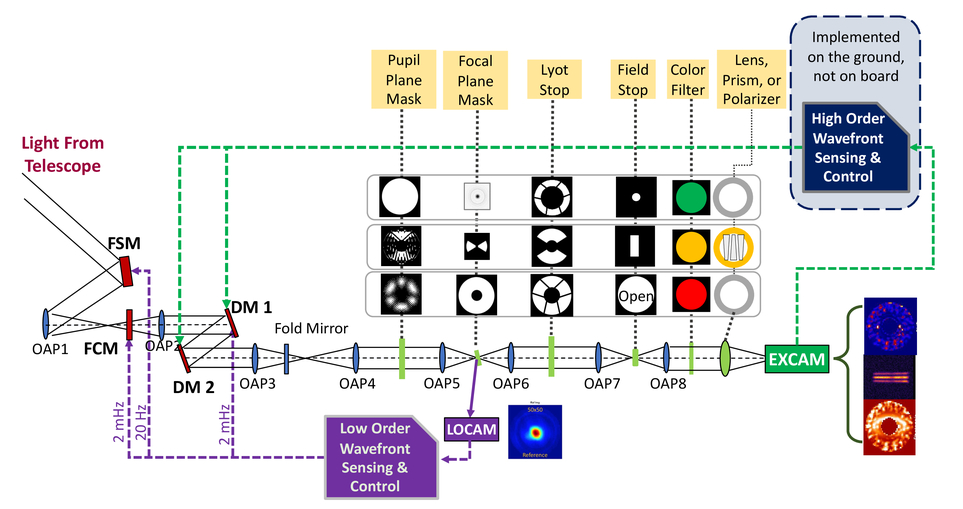}
	\caption{\label{fig:hlc_model} The CGI optical train and wavefront sensing and control architecture. The optical elements of the HLC mode of interest are depicted in the top panel. Before an observation, the high order wavefront sensing and control loop is performed on a bright reference star to generate a `dark hole' (an area where starlight is suppressed). Then, the DM shapes are fixed, and the telescope slews to the target star for the observation. During the observation, wavefront errors accrue as a result of instrumental disturbances and drifts, the effects of which this work aims to mitigate in post-processing.  Figure from \citet{kasdin_2020}.}
\end{center}
\end{figure*}

In this section, the use of robust observables with the Hybrid Lyot Coronagraph of the Roman Space Telescope is analyzed through simulation. However, this approach could also be applied to other coronagraphs, as long as the exposure times are short enough such that wavefront error has not been averaged out. The optical model of CGI is shown in Figure \ref{fig:hlc_model} \citep{kasdin_2020}. The optical elements corresponding to the HLC mode (the relevant mode for this work) are depicted in the top panel.

\subsection{Optical Model} \label{sec:hlc_model}
The HLC operates around a dark hole state, which is obtained using focal plane wavefront control (with deformable mirrors) to measure and minimize the electric field in the detector plane. Such focal plane wavefront control significantly suppresses the amount of starlight in the dark hole, and allows for much deeper raw contrasts than with just a flattened wavefront. Before an observation, the dark hole is generated using high order wavefront sensing and control loop on a bright reference star. Then, the DM shapes are fixed, and the telescope slews to the target star for the observation. During the observation, wavefront errors accrue as a result of instrumental disturbances and drifts. This work aims to mitigate the effects of those wavefront errors in post-processing. Note that as a result of the dark-hole generation, the nominal electric field $\bm{E_{s0}}$ is not a flat wavefront, but the pupil plane electric-field obtained at the end of the dark-hole generation sequence.

A Lightweight Space Coronagraph Simulator (LSCS) \footnote{https://github.com/leonidprinceton/LightweightSpaceCoronagraphSimulator} derived from the HLC model in the Fast Linearized Coronagraph Optimizer \citep[FALCO;][]{riggs_falco} toolbox is used for the following simulations. The LSCS relies on the HLC numerical model and focal-plane wavefront control algorithm included in FALCO to first generate the initial dark hole electric field. The numerical model in FALCO is also used to calculate $\mathbf{C}$ from the finite-difference sensitivities of the focal plane electric field to pupil plane phase error expressed in the Zernike basis (we have made the assumption that the matrix transformation is approximately linear in phase, valid when the phase error is much less than a wave). Although we use finite-differences to calculate $\mathbf{C}$, one could also construct it by multiplying together all the matrix transformations of the optical model. These simulations are conducted at a single wavelength of 546 nm.

The average raw contrast of the initial dark hole is $5.6\times10^{-9}$. The LSCS model takes in Zernike coefficients for phase aberrations, calculates their effect on the focal plane electric field, and adds them to the initial dark hole electric field to obtain the focal plane electric field in the presence of wavefront errors. The intensity can be calculated as the norm-square of the focal plane electric field. Detector and photon noise are not simulated. Since the default LSCS models only the first 136 Zernikes, FALCO is first used to extend the LSCS model to 528 Zernikes in order for the entire dark hole to be sampled. 

This results in using 528 Zernikes to sample the entire dark hole, or a $N_\text{basis}$ of 528. The LSCS models a detector that is $74 \times 74$ pixels, with 3 pixels per $\lambda/D$, for a total pixel number of $N_\text{pix} = 5476$. The number of pixels defined to be in the dark hole is $N_{\text{DH}}=2608$. This model does not consider the effects of amplitude errors, and only analyzes phase errors, which, from end-to-end modeling of Roman CGI, are expected to be the dominant form of dynamic aberrations \citep{krist_2023}. However, for a system where dynamic amplitude errors are comparable to dynamic phase errors, both should be included.

\subsection{Response Matrices} \label{sec:hlc_response_matrices}
The Zernike coefficient drift values from the Observing Scenario simulations \citep[OS 9;][]{krist_os9}, based on physical modeling of the telescope, indicate that the WFE expected on Roman will fall within the linear regime of this dark hole. However, the level of wavefront error may end up being higher than currently expected. Additionally, on ground based telescopes, wavefront error from adaptive optics residuals is typically in the quadratic regime. Therefore, for illustrative purposes, both a linearly-dominated noise model and quadratically-dominated noise model are examined.

The matrices $\mathbf{A_l}$ and $\mathbf{A_q}$ are calculated according to Section \ref{sec:response_matrices}. The linear matrix includes all Zernikes present in the optical model, and thus has an input dimension of $N_{\text{basis}} = 528$. The quadratic matrix includes only the first 100 Zernikes, and thus has an input dimension of ${N_{\text{redu}}+1 \choose 2} = 5,050$. The relevant dimensions of the objects used in this analysis are listed in Table \ref{tab:hlc_dimensions}. Note that the cutoff number $N_m$ is a variable to optimized over.

\begin{table}[h!]
  \begin{center}
    \caption{Quantities and Dimensions for Analysis of the Roman Space Telescope HLC}
    \begin{tabular}{c|c|c|c}
      \textbf{Quantity}  & \textbf{Description} & \textbf{Dimension (Dependency)} & \textbf{Dimension (Value)} \\
      \hline
      $\bm{\bm{E_{\text{pup}}}}$ & Vector of electric in pupil plane
      & $N_\text{basis}$ & $528$\\
      $\bm{\bm{E_\text{det}}}$ & Vector of detector plane electric field
      & $N_\text{pix}$ & $5,476$\\
      $\bm{I_{\text{det}}}$ & Vector of detector plane intensity
      & $N_\text{pix}$ & $5,476$ \\
      $\bm{E_{\text{DH}}}$ & Vector of detector plane electric field in dark hole
      & $N_{\text{DH}}$ & $2,608$\\
      $\bm{I_{\text{DH}}}$ & Vector of detector plane intensity in dark hole
      & $N_{\text{DH}}$ & $2,608$\\
      $\mathbf{A_l}$ & Linear-regime instrument response matrix
      & $N_\text{pix} \times N_\text{basis}$ & $5,476 \times 528$\\
      $\mathbf{U_l}$ & Left singular matrix of $\mathbf{A_l}$
      & $N_\text{pix} \times N_\text{pix}$ & $5,476 \times 5,476$\\
      $\mathbf{S_l}$& Singular value matrix of $\mathbf{A_l}$ & $N_\text{pix} \times N_\text{basis}$ & $5,476 \times 528$\\
      $\mathbf{V_l}$ & Right singular matrix of $\mathbf{A_l}$ 
      & $N_\text{basis} \times N_\text{basis}$ & $528 \times 528$\\
      $\mathbf{K_l}$ & Linear-regime projection matrix
      & $(N_\text{pix}-N_{m}) \times N_{\text{DH}}$ & $(N_\text{pix}-N_{m}) \times 2,608$\\
      $\bm{O_l}$ & Vector of linear-regime observables
      & $(N_\text{pix}-N_{m})$ & $(N_\text{pix}-N_{m})$\\
      $\mathbf{A_q}$ & Quadratic-regime instrument response matrix & $N_\text{pix} \times {N_{\text{redu}}+1 \choose 2}$ & $5,476 \times 5,050$\\
      $\mathbf{U_q}$ & Left singular matrix of $\mathbf{A_q}$ & $N_\text{pix} \times N_\text{pix}$ & $5,476 \times 5,476$\\
      $\mathbf{S_q}$& Singular value matrix of $\mathbf{A_q}$ & $N_\text{pix} \times {N_{\text{redu}}+1 \choose 2}$ & $5,476 \times 5,050$\\
      $\mathbf{V_q}$ & Right singular matrix of $\mathbf{A_q}$ 
      & ${N_{\text{redu}}+1 \choose 2} \times {N_{\text{redu}}+1 \choose 2}$ & $5,050 \times 5,050$\\
      $\mathbf{K_q}$ & Quadratic-regime projection matrix
      & $(N_\text{pix}-N_{m}) \times N_{\text{DH}}$ & $(N_\text{pix}-N_{m}) \times 2,608$\\
      $\bm{O_q}$ & Vector of quadratic-regime observables
      & $(N_\text{pix}-N_{m})$ & $(N_\text{pix}-N_{m})$
     
    \end{tabular}
  \end{center}
  \label{tab:hlc_dimensions}
\end{table}

\newpage
\subsection{Projection Matrices} \label{sec:projection_matrices}
\begin{figure*}[!ht]
\begin{center}
	\includegraphics[scale = 0.6]{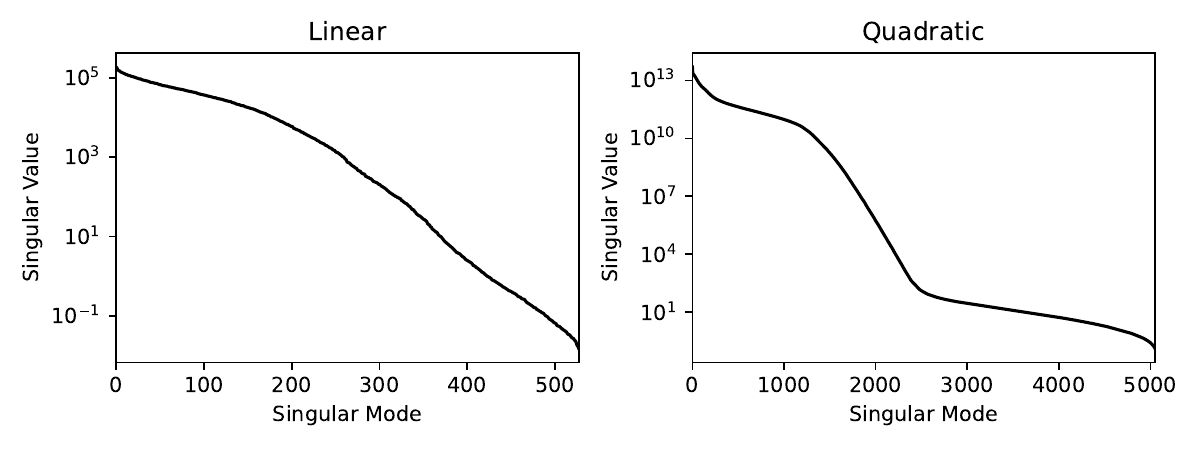}
	\caption{\label{fig:hlc_svd} The singular values of $\mathbf{A_l}$ (left) and $\mathbf{A_q}$ (right). Note that the transfer matrices are rectangular and have $N_\text{pix} = 5476$ total singular modes, but the singular values beyond the size of the input dimension are all 0.}
\end{center}
\end{figure*}

\begin{figure*}[!ht]
\begin{center}
    Singular Modes of $\mathbf{A_l}$
    
	\includegraphics[scale = 0.8]{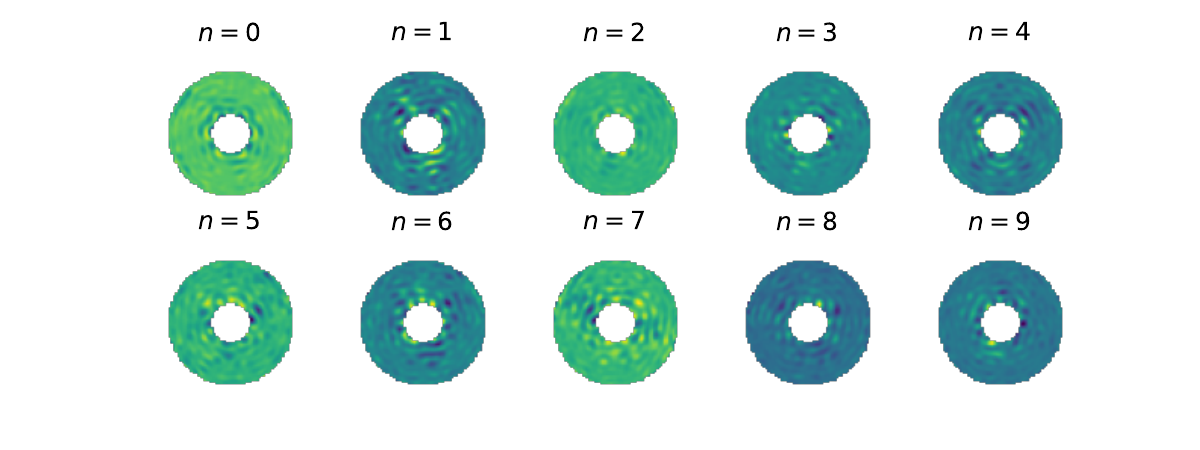}
	
	Singular Modes of $\mathbf{A_q}$
	
	\includegraphics[scale = 0.8]{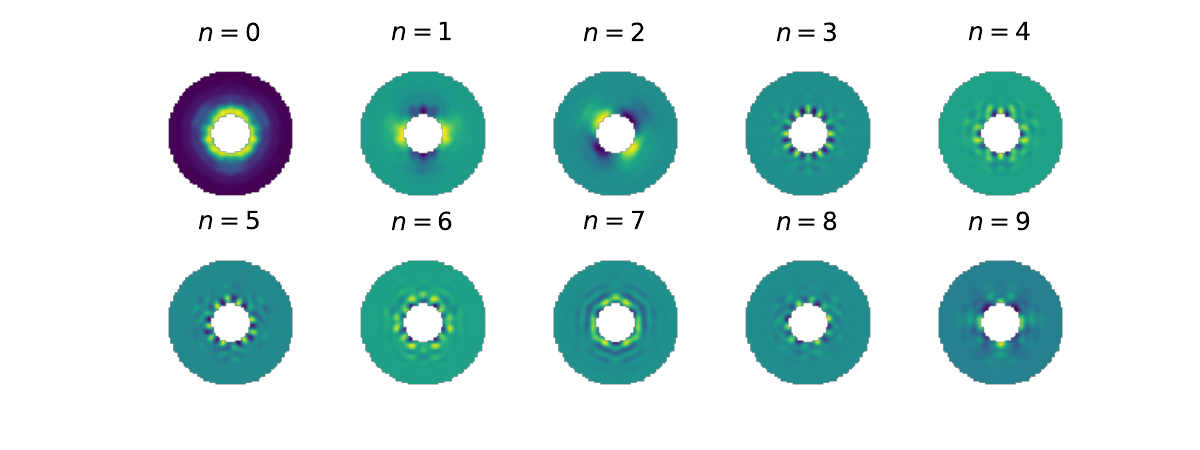}
	\caption{\label{fig:hlc_modes} Top: The first 10 singular modes of $\mathbf{A_l}$ as represented in the the detector plane intensity basis (linear scale). Bottom: The first 10 singular modes of $\mathbf{A_q}$ as represented in the the detector plane intensity basis (linear scale). The HLC design is nearly circularly symmetric, broken only by the six secondary mirror struts (which can also be seen in the Lyot stop). Because the quadratic transfer matrix depends only on the coronagraph operator $\mathbf{C}$, its singular modes exhibit cosine and sine-like azimuthal behavior associated with circularly symmetric operators. However, the linear transfer matrix depends on both $\mathbf{C}$ as well as on the focal-plane electric field at the end of dark hole creation, which is random and not circularly symmetric. Thus, its singular modes show no such symmetry structures. These singular modes correspond to the intensity patterns most likely to be attributed to wavefront error. Meanwhile, the companion's intensity pattern (the PSF at its location) overlaps very little with these dominant modes, so its signal is mostly retained when the dominant modes are projected out.}
\end{center}
\end{figure*}

According to Algorithm \ref{alg:calc_K}, a singular value decomposition of $\mathbf{A}=\mathbf{USV}^T$ is performed for each transfer matrix, revealing their singular modes and corresponding singular values. The singular values of the transfer matrices are shown in the top of Figure \ref{fig:hlc_svd}. The first 10 singular modes of each transfer matrix as represented in the detector plane intensity basis (with pixels not in the dark hole masked) are plotted in Figure \ref{fig:hlc_modes}.

From both the linear and the quadratic transfer matrix, model-based projection matrices with a range of cutoff modes are calculated according to Algorithm \ref{alg:calc_K}. To rule out the effect of dimensionality alone on the dataset, random projection matrices of the same size are also generated. This is done by taking the SVD of a matrix the same size as the $\mathbf{A}$ matrices, but populated with values drawn uniformly from -1 to 1, and then removing the same number of dominant modes as is done with $\mathbf{A}$. These matrices are applied to synthetically generated data to quantify their effect on the detectability of binary companion signals.

\subsection{Synthetic Data Analysis} \label{sec:syndata_analysis}

\subsubsection{Synthetic Data Generation} \label{sec:syndata_gen}

FALCO is used to generate a library of off-axis PSFs corresponding to the dark hole state, which can be injected as binary companions. These off-axis PSFs do not incorporate any WFE that is added on top of the dark hole state. However, the effect of WFE on the off-axis signal is expected to be much, much smaller than its effect on the on-axis stellar signal, so not modeling the effects of wavefront error on the off-axis signal should have a negligible impact on the data.

The optical system is first initialized in the dark hole state. Two noise models are considered, one in the linear regime, and one in the quadratic regime. Each dataset thus consists of 20 instantaneous frames of independent noise realizations. For each frame, the spatial PSD given in Equation \ref{eqn:PSD_hlc} is used to generate the wavefront error.

\begin{equation}
\label{eqn:PSD_hlc}
PSD(n_z) = a{n_z}^{b}
\end{equation}

In this equation, $n_z$ is Noll-ordered index of the Zernike coefficient. The normalization parameter $a$ is chosen to be 10 nm for the linear regime, and 130 nm for the quadratic regime. The power law exponent $b$ is chosen to be -2. These PSDs correspond to an average wavefront error (calculated over 100 realizations) of about 7 nm (0.013 waves) RMS for the linear regime data, and about 110 nm (0.2 waves) RMS for the quadratic regime data. As discussed in Section \ref{sec:data_formation}, the linear-quadratic transition occurs at approximately 0.1 waves RMS. Although 110 nm RMS of dynamic wavefront error is unrealistically high for the Roman HLC, we include this regime for demonstration purposes, as this level of WFE would be relevant on ground-based telescopes.

\begin{figure*}[!ht]
\begin{center}
	\includegraphics[scale = 0.7]{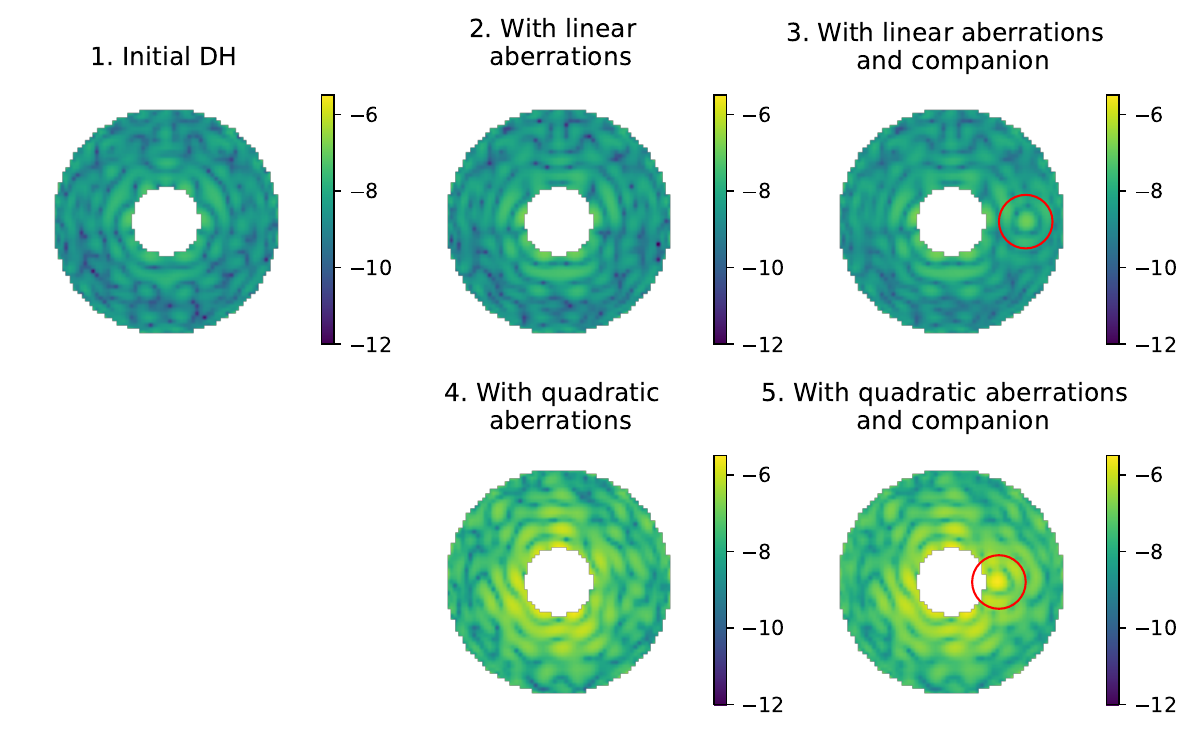}
	\caption{\label{fig:hlc_syndata} 1) Initial dark hole intensity achieved using electric field conjugation with the HLC. 2) A single snapshot with linear-regime wavefront aberrations. 3) The same snapshot with an injected companion with a flux ratio of $2 \times 10^{-7}$ at $6.5\,\lambda/D$ (indicated with red circle). 4) A single snapshot with quadratic-regime wavefront aberrations. 5) The same snapshot with an injected companion with a flux ratio of $5 \times 10^{-6}$ at $4\,\lambda/D$ (indicated with red circle). All intensities are shown in $\text{log}_{10}$ of raw contrast.}
\end{center}
\end{figure*}

The resulting 528 Zernike coefficients are propagated through the LSCS to calculate the resulting dark hole intensities. In order to create data with an injected companion planet, the off-axis PSF at the desired separation is scaled by the companion's flux ratio, and then added to the dark hole intensity. The separation of the injected companion is set to be $6.5\,\lambda/D$ in the linear case (which is the middle of the dark hole) and $4.0\,\lambda/D$ in the quadratic case (since the model is only valid within $\sim 5\,\lambda/D$). The position angles of both are set to be 0. Frames without the injected companion are used for the control case. Figure \ref{fig:hlc_syndata} shows example data frames: the initial dark hole, example frames with the aberrations from both noise models applied, and the same frames with injected companion signals. The flux ratio of the companion is $2 \times 10^{-7}$ for frame with linear-regime errors and $5 \times 10^{-6}$ for the frame with quadratic-regime errors. These flux ratios correspond to particularly bright planets chosen to be visible by eye.

\begin{figure*}[!ht]
\begin{center}
	\includegraphics[scale = 0.8]{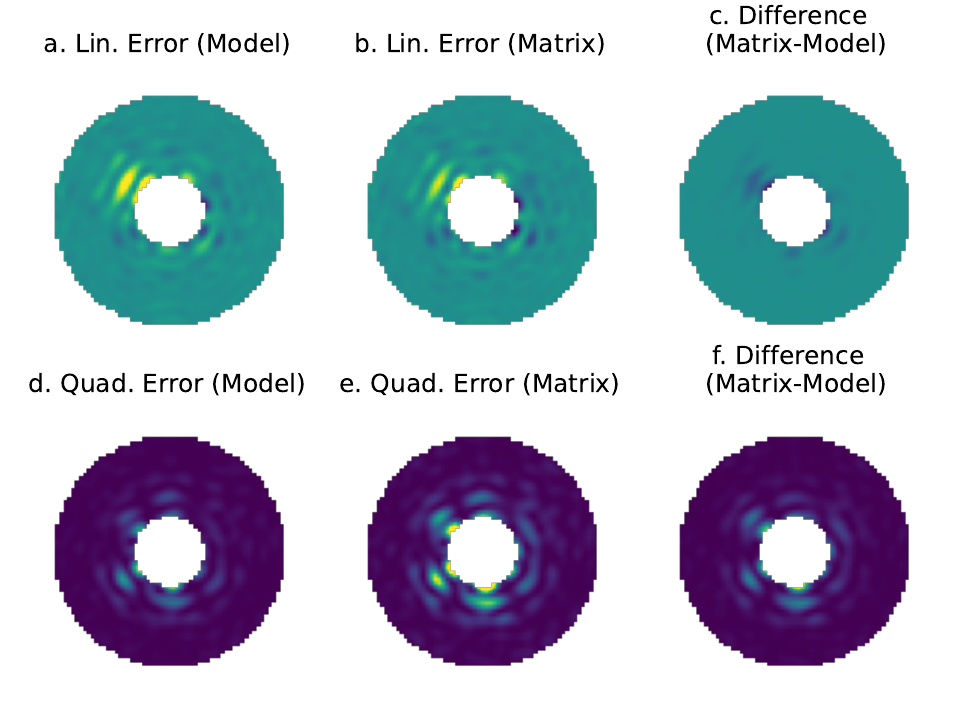}
	\caption{\label{fig:hlc_error_comparison} a. Example linear-regime intensity error from the optical model. b. Corresponding linear-regime intensity error reconstructed by response matrix $\mathbf{A_l}$, plotted on the same scale as (a). c. The difference between the response matrix prediction and the optical model prediction, plotted on the same scale as (a) and (b). d. Example quadratic-regime intensity error from the optical model. e. Corresponding quadratic-regime intensity error reconstructed by the response matrix $\mathbf{A_q}$, plotted on the same scale as (d). f. The difference between the response matrix prediction and the optical model prediction, plotted on the same scale as (d) and (e). Slight differences arise because the model includes both the linear and quadratic error terms while the matrix predictions only include one or the other, i.e. the linear matrix prediction neglects the contribution of the quadratic term and the quadratic matrix prediction neglects the contribution of the linear term (as well as the influence of any Zernikes past the first 100). While the linear matrix prediction is biased low near the peaks and the quadratic matrix prediction biased high overall, our method depends only on how well the \textit{spatial structure} of the errors are reproduced. A relevant metric for characterizing the spatial overlap is the normalized inner product between the optical model prediction and the transfer matrix prediction, where a value of 1 indicates perfect spatial overlap and a value of 0 indicates perfect spatial orthogonality. In this case, the normalized inner product is 0.936 for the linear regime example and 0.985 for the quadratic regime example, sufficient for providing a quantifiable improvement in detection sensitivity.}
\end{center}
\end{figure*}

It is worthwhile to examine how well the response matrices calculated in Section \ref{sec:hlc_response_matrices} can reconstruct the intensity errors present in the synthetic data. Figure \ref{fig:hlc_error_comparison} compares the intensity error resulting from WFE as calculated from the optical model with the intensity error calculated by multiplying the WFE by the appropriate response matrix, for example frames in both the linear and the quadratic regimes. In both regimes, the response matrices largely reproduce the spatial structure of the intensity error from the optical model.

\subsubsection{Processing Synthetic Data} \label{hlc_syndata_analysis}
The quantity $|\mathbf{C} \bm{E_{s0}}|^2$ is  the initial dark hole intensity without any extra WFE applied (as determined from the data at the end of the dark-hole digging sequence on the reference star, for example). This nominal signal is first subtracted from each frame. Then, the pixels within the defined dark hole are gathered into the vector $\bm{\mathit{\Delta} I_{\text{DH}}}$. The data is left-multiplied by the appropriate $\mathbf{K}$ matrix to obtain the observables $\bm{O} = \mathbf{K} \bm{\mathit{\Delta} I_{\text{DH}}}$. The data is also left-multiplied by the random matrix of the same size as $\mathbf{K}$ to obtain data whose dimension has been reduced randomly. For each case, the average of the data over the twenty frames is used as the final measurement, while the standard deviation of the frames is used as the measurement uncertainty. Note that the process outlined does not rely on reference stars or dithering by deformable mirrors, and can be used even on observations for which reference observations or wavefront diversity is unavailable.

\subsubsection{Flux Ratio Detection Limits}
\begin{figure*}[!ht]
\begin{center}
	\includegraphics[scale = 0.33 ]{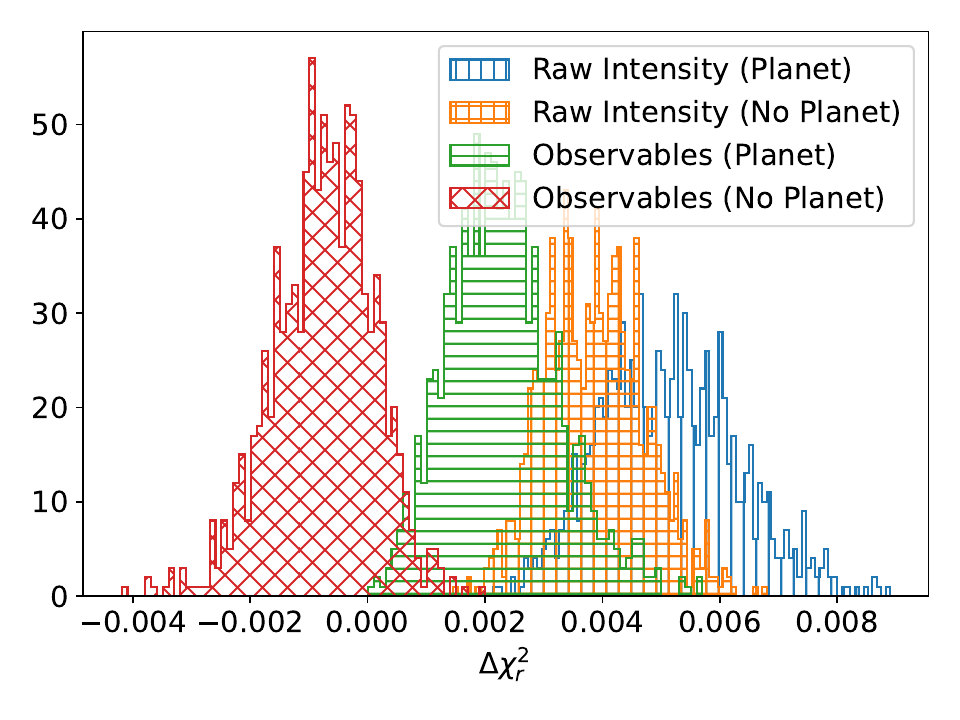}
	\includegraphics[scale = 0.33 ]{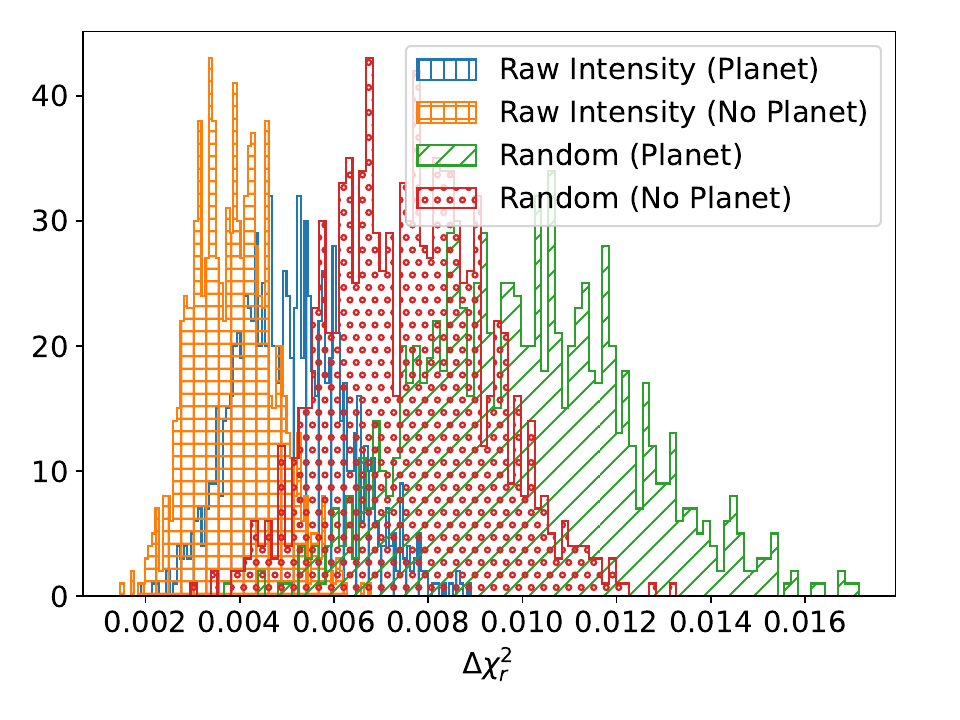}
	\includegraphics[scale = 0.33 ]{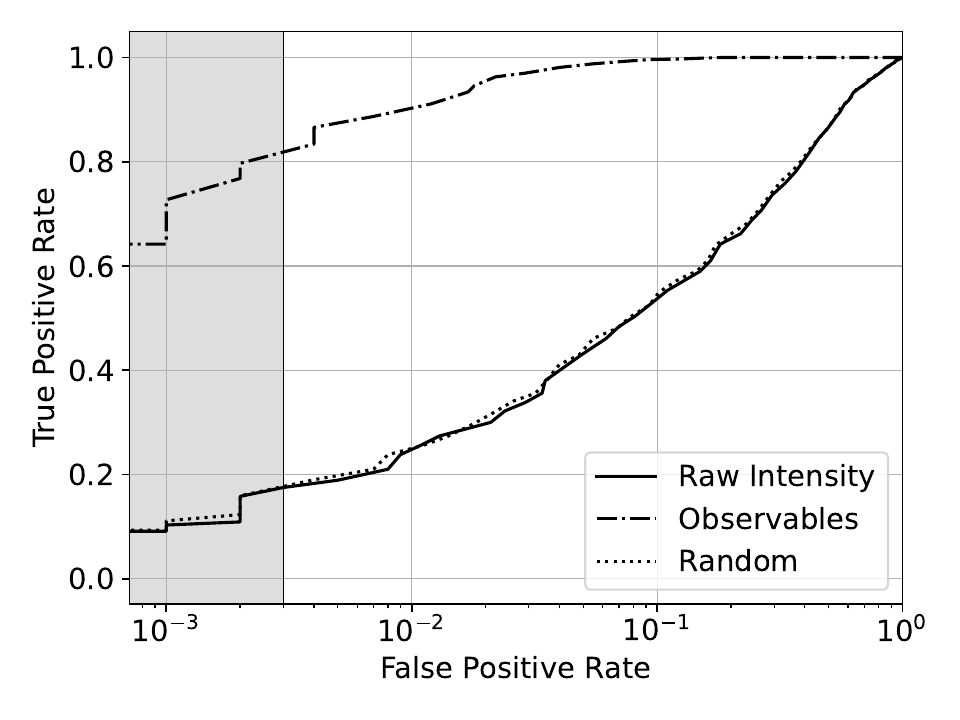}
	\caption{\label{fig:hlc_hists} Detection test results for the quadratic regime noise model. The companion planet considered has flux ratio of $5.4 \times 10^{-7}$ and is located at $4.0\,\lambda/D$. Left: Histograms from using raw intensities compared to those from using quadratic robust observables with the optimal cutoff of $N_m=70$. The histograms using raw intensity overlap significantly, making it difficult to distinguish between a model with a planet and a model without one, while the histograms using the robust observables are further separated and more distinguishable. Middle: Histograms for using raw intensities and a random projection matrix of the same size as the instrumentally-motivated projection. Both sets of histograms overlap significantly, and the random projection does not improve the distinguishability of the two models. Right: ROC curves corresponding to the histograms. Grey area indicates false positive rates which are not well sampled as they involve less than 3 datasets with false detections. The ROC curve shows that while using the robust observables results in a $\mathrm{FPR}=0.01$ and $\mathrm{TPR}=0.9$ detection of the injected planet, both the raw intensity and the randomly dimensionally reduced data remain very far from detectability.
}
\end{center}
\end{figure*}

\begin{figure*}[!ht]
\begin{center}
	\includegraphics[scale = 0.5 ]{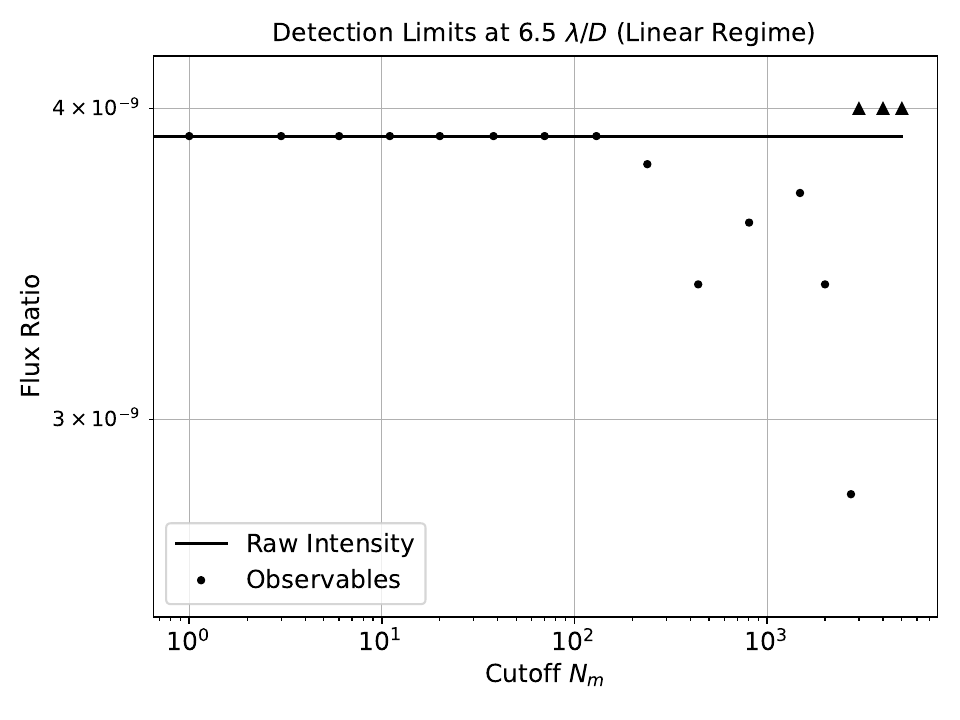}
	\includegraphics[scale = 0.5 ]{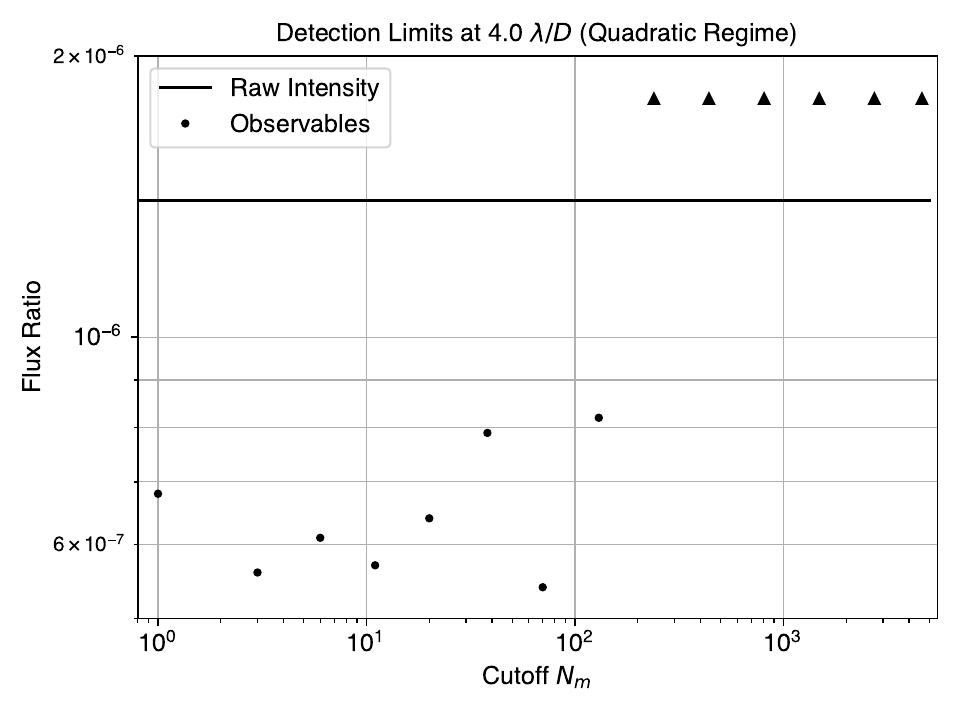}
	\caption{\label{fig:hlc_contrast_curves} Flux ratio detection limits ($\mathrm{FPR}=0.01$, $\mathrm{TPR}=0.90$) for a binary companion (to two significant figures) as a function of cutoff mode. Upward triangles indicate that a projection matrix with the specified cutoff mode performs worse than using the raw intensity, which occurs when the modes the majority of the planet signal overlaps with have also been projected out. Left: Linear regime with a companion at $6.5 \,\ \lambda/D$. The optimal cutoff mode is 2,727, which results in a detection limit of $2.8\times 10^{-9}$. Unshowable in log-log scale is the detection limit with $N_m=0$, which, with observables, is $3.9 \times 10^{-9}$. This is, as expected from the fact that no error modes are removed, equal to the raw intensity detection limit. Right: Left: Quadratic regime with a companion at $4.0\,\lambda/D$. The optimal cutoff mode is 70, which results in a detection limit of $5.4 \times 10^{-7}$. Unshowable in log-log scale is the detection limit with $N_m=0$, which, with observables, is $1.4 \times 10^{-6}$. This is, as expected from the fact that no error modes are removed, equal to the raw intensity detection limit.}
\end{center}
\end{figure*}

Detection tests are applied to these measured intensities and observables in order to characterize the detectability of a companion with these measurements. Detection limits are determined using the Monte Carlo method. One thousand random datasets are generated for each noise model with a given flux ratio. Each dataset is processed as raw intensity data, and with each projection matrix with a different cutoff mode, and the $\mathit{\Delta} \chi^2_r$ values are calculated for each case. Figure \ref{fig:hlc_hists} shows example histograms of the resulting $\mathit{\Delta} \chi^2_r$ values for a $c=5.4 \times 10^{-7}$ at $4.0\,\lambda/D$ companion with the quadratic noise model, as well as the corresponding ROC curves, for the projection matrix with cutoff mode $N_m=70$ (which, as shown in Figure \ref{fig:hlc_contrast_curves}, is the optimal cutoff at this spatial separation). The ROC curve shows that while using the robust observables results in a $\mathrm{FPR}=0.01$ and $\mathrm{TPR}=0.9$ detection of the injected companion, both the raw intensity and the randomly dimensionally reduced data remain very far from detectability.

This process is repeated for a range of flux ratios (to a precision of two significant figures). The resulting $\mathrm{FPR}=0.01$ and $\mathrm{TPR}=0.9$ detection limits for both regimes, as a function of cutoff mode $N_m$, are shown in Figure \ref{fig:hlc_contrast_curves}. Note that these flux ratio detection limits are not based on any statistical assumptions or extrapolations, but rather real FPRs and TPRs calculated by analyzing one thousand synthetically generated datasets, with injected companions of the given flux ratios and separations. The results show that with the linear-regime noise model, the robust observables increases the detectability of a companion at $6.5\,\lambda/D$ by 28\%. With the quadratic-regime noise model, using robust observables increases the detectability of a companion at $4.0\,\lambda/D$ by over a factor of two, and the improvement is not particularly sensitive to $N_m$ beyond the first few modes. For the linear regime, this approach can also easily be extended to companions throughout the entire dark hole, though significant computation would be required to optimize $N_m$ at all separations. For the quadratic regime, our model is only valid within $\sim 5\,\lambda/D$, though Appendix \ref{app:a} discusses a method that can be used to extend the spatial coverage without incurring impractical computational costs.

\section{Discussion}

\subsection{Temporally Correlated WFE and Compatibility with Other Post-processing Techniques}

This work aims to characterize the effect of using robust observables in isolation. Thus, only noise models in which the WFE is uncorrelated in time are examined, since additional post-processing techniques are typically used to handle time-correlated data. Robust observables are compatible with these other post-processing techniques, and can serve as an instrument-motivated prior in the overall post-processing strategy. For example, random errors can first be reduced by projecting the data into a subspace that is robust to wavefront error. Then, reference observations along with PCA-based methods such as KLIP \citep{soummer_klip} can be used to calibrate static and quasi-static errors and de-correlate the frames in time. This is similar to the calibration approach used in non-redundant aperture masking (NRM) interferometry or kernel-phase interferometry, in which data is projected onto closure-phases or kernel-phases respectively, which are then calibrated based on reference observations \citep{martinache2010,ireland_2013,pope2021}. A more sophisticated approach would be to formulate post-processing as a statistical inference problem, where a least-squares fit with the reference frames makes one up term in the cost function, and a prior over the instrumental modes (e.g. weighted by the singular value spectrum) makes up another term.

\citet{ygouf_2016} shows that for the time-varying wavefront error expected on the Roman Space Telescope HLC, classical PSF subtraction with a reference observation increases the contrast gain by a factor of a few to about ten, depending on the scenario. Future work includes investigating how much overall post-processing gain can be achieved when robust observables and calibration strategies are combined, and which hybrid strategies maximize the sensitivity that can be obtained with all available information.

\subsection{PSD Engineering}

The robust observables derived in this work are agnostic to the actual temporal or spatial PSD of the static and dynamical wavefront errors, and are intended to be applied when these PSDs are not well-known or imperfectly characterized. As of today, this is the case for all ground-based instruments (as predictions of the influence of the atmosphere are quite imperfect), and space-based  missions (as HST and JWST observatory level key metrics for requirements are expressed in terms of encircled energy, not contrast). However, it has been proposed that for future space telescope coronagraphs, the telescope WFE PSD must comply with stringent requirements in order to facilitate exoplanet detection \citep{nemati_2020}.

For instance, the PASTIS approach \citep{leboulleux2018pairbased, laginja_2019} considers the effects of the quadratic response on the \textit{average} intensity contrast over the entire dark hole (or region of interest), calculating which modes the coronagraph is most sensitive to in order to determine stability tolerances for the segments accordingly. Calculating robust observables for post-processing is akin to doing PASTIS backwards, where the modes the coronagraph is most sensitive to are calculated in order to project them out of the data. For such telescopes, that have PSDs engineered based on the instrument response, the additional gain from using robust observables will depend on how well the error modes are suppressed in hardware, as well as the timescales at which power in those modes leaks through. To some extent, robust observables will remain applicable to such future telescopes and instruments in the spatial and temporal sub-spaces in which they do not meet their requirements.

\subsection{Model Accuracy}
In this analysis, the model used to generate the instrument response matrices is exactly the same model that is used to generate the synthetic data. In a real observations, the instrument model will not exactly match the behavior of the actual instrument, and one future avenue to explore is how well a model must match the instrument in order for robust observables to work on real data. This technique's robustness can be investigated by first calculating the response matrices using one model, then changing the parameters of the model (e.g. the coronagraphic mask size and displacement, the DM alignment, the detector pixel scale) before generating synthetic data, and examining how well the robust observables work in the presence of model mismatch.

For instruments equipped with wavefront modulating devices such as deformable mirrors, however, the instrument response matrix may also be calculated experimentally. If a perturbation within the linear regime is applied, the difference in measured intensity can be directly registered into the appropriate column of the linear response matrix. The technique for experimentally building the quadratic response matrices is equivalent to the approach used for PASTIS  \citep{laginja_2019}, with the difference that the measurements are not averaged over a dark hole, but rather maintained for every pixel. Additionally, some wavefront and control schemes such as implicit electric-field conjugation \citep{haffert_2023} already involve an empirical measurement of the instrument response, which can be used to derive linear-regime robust observables without having to set aside additional calibration time. Experimentally building instrument response matrices circumvents the need to have a well-matched numerical model, and allows for the response matrices to capture effects in the real instrument.
 
\section{Conclusions}

A coronagraph model with linear and quadratic contributions of wavefront error to detector plane intensity is developed, and when either term is dominant, the coronagraph response can be approximated by a transfer matrix. A useful projection can be found from this transfer matrix that removes the dominant error modes, resulting in observables that are more robust to WFE in the regime of interest. These robust observables are extracted from synthetically generated data with the Hybrid Lyot Coronagraph of the Roman Space Telescope in both the linear and quadratic regimes. The performance of the robust observables is compared to that of the raw intensity data through calculations of their respective binary companion flux ratio detection limits. In these examples, using the robust observables significantly increases the sensitivity to the signal of a binary companion. A projection onto a robust subspace can in theory be combined with other families of post-processing algorithms. Hybrid post-processing approaches would incorporate information on the instrument response alongside the other available information (such as angular diversity, spectral diversity, reference observations, or WFC telemetry) to fully maximize the sensitivity to astrophysical signals in coronagraphic data; however, the approach outlined in this work can be applied to observational data and result in post-processing gains even if such additional information is unavailable.

\acknowledgments


We thank the anonymous reviewer for their careful consideration and feedback. We thank Frantz Martinache, Mamadou N'Diaye, and Alban Ceau for helpful discussions on this topic, and to Dimitri Mawet for his additional perspectives and support. 

This work is supported by the National Science Foundation Graduate Research Fellowship under Grant No. 1122374.


This research made use of NASA's Astrophysics Data System.

BJSP would like to acknowledge the traditional owners of the land on which the University of Queensland is situated, the Turrbal and Jagera people. We pay respects to their Ancestors and descendants, who continue cultural and spiritual connections to Country. 

%

\vspace{5mm}


\software{This research made use of FALCO, the Fast Linearized Coronagraph Optimizer \citep{riggs_falco}; the Lightweight Space Coronagraph Simulator (\href{https://github.com/leonidprinceton/LightweightSpaceCoronagraphSimulator}{https://github.com/leonidprinceton/LightweightSpaceCoronagraphSimulator}); Astropy \citep{astropy:2013, astropy:2018}; NumPy \citep{harris2020array}; SciPy \citep{2020SciPy-NMeth}; and Matplotlib \citep{Hunter:2007_matplotlib}.}



\appendix

\section{Quadratic Model Approximation and Extension} \label{app:a}

As explained in Section \ref{finding_k}, the calculation of the projection matrix involves a singular value decomposition (SVD) of the response matrix, but a quadratic response matrix that includes all 528 Zernikes needed to span the dark hole would have a size of $5,476 \times 139,656$. Since calculating the SVD of a matrix of this size is too computationally burdensome, we explore an approximation of the quadratic response that models only the impact of norm-squared of each input basis vector while neglecting the effects of the pairwise combinations. Namely, we use an approximate response matrix $\mathbf{A_q'}$ with elements:

\begin{equation} \label{eq:A_q_prime}
    \mathrm{A_q'}_{kj} = \mathrm{C}_{kj}^*\mathrm{C}_{kj}.
\end{equation}

The index $j$ labels the input basis vector and the index $k$ labels the detector pixel. The size of $\mathbf{A_q'}$ scales linearly with the number of Zernike models, and in our case would be of size $5,476 \times 528$, which is easily decomposable.

\begin{figure*}[!ht]
\begin{center}
	\includegraphics[scale = 0.75]{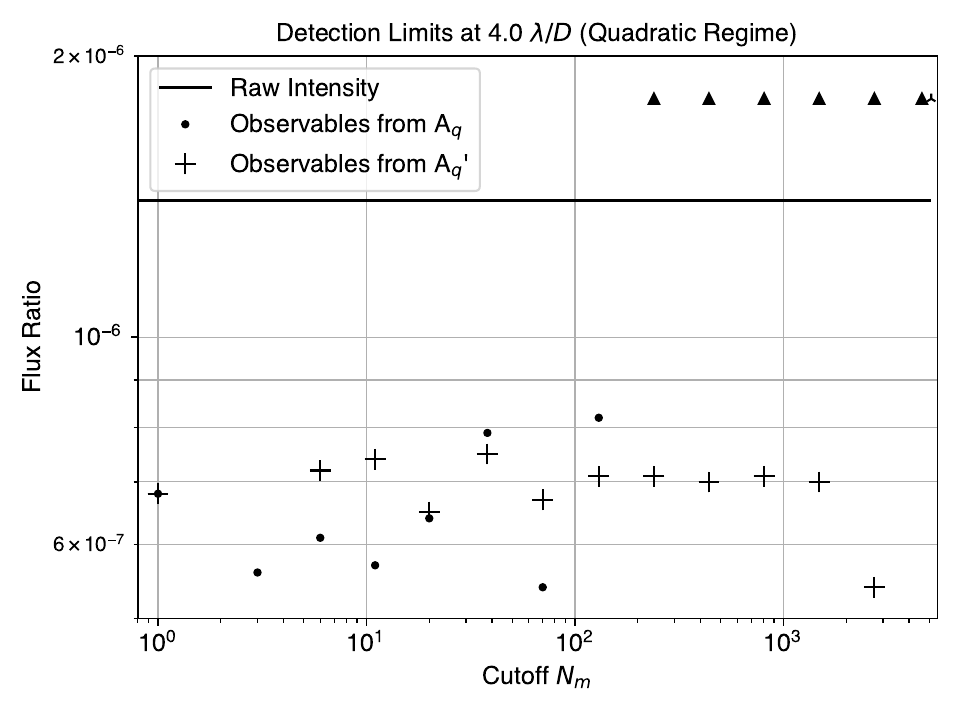}
	\caption{\label{fig:fr_limits_reduced} Quadratic regime flux ratio detection limits ($\mathrm{FPR}=0.01$, $\mathrm{TPR}=0.90$), to two significant figures, as a function of cutoff mode, for a companion at $4.0\,\ \lambda/D$. Only the first 100 Zernikes are used in the model used to calculated the full and approximate quadratic transfer matrices, but WFE up to 538 Zernikes are included in the synthetic data. Upward triangles or spikes indicate that a projection matrix with the specified cutoff mode performs  worse than using the raw intensity, which occurs when the modes the majority of the planet signal overlaps with have also been projected out. The full matrix achieves the best results with a cutoff mode of 70, leading to a detection limit of $5.4\times10^{-7}$ while the approximate matrix achieves the best results with a cutoff mode of 2,727, also leading to a detection limit of $5.4\times10^{-7}$. Unshowable in log-log scale is the detection limit with $N_m=0$, which, with observables, is $1.4 \times 10^{-6}$. This is, as expected from the fact that no error modes are removed, the same as the raw intensity detection limit of $1.4 \times 10^{-6}$. These results indicate that the approximation performs as well as the full model.}
\end{center}
\end{figure*}

Note that $\mathbf{A_q'}$ cannot be used to accurately reproduce quadratic-regime intensity error. However, $\mathbf{A_q'}$ is nevertheless useful for identifying a subspace robust to quadratic regime wavefront errors, leading to increased signal-to-noise. We can observe this by comparing the detection test results with and without using the approximation for a model with 100 Zernikes. We calculate the approximation $\mathbf{A_q'}$ using Equation \ref{eq:A_q_prime}, and use the original $\mathbf{A_q}$ from Section \ref{sec:projection_matrices}. Detection tests on quadratic-regime synthetic data, similar to the one from Section \ref{sec:syndata_analysis} are performed, using projection matrices derived from both $\mathbf{A_q}$ and $\mathbf{A_q'}$. The resulting flux ratio detection limits as a function of cutoff mode are shown in Figure \ref{fig:fr_limits_reduced}.

The full matrix achieves the best results with a cutoff mode of 70, leading to a detection limit of $5.4 \times10^{-7}$ while the approximate matrix achieves the best results with a cutoff mode of 2,727, also leading to a detection limit of $5.4\times10^{-7}$. These results show that the approximation performs as well as the full model.

\begin{figure*}[!ht]
\begin{center}
	\includegraphics[scale = 0.8]{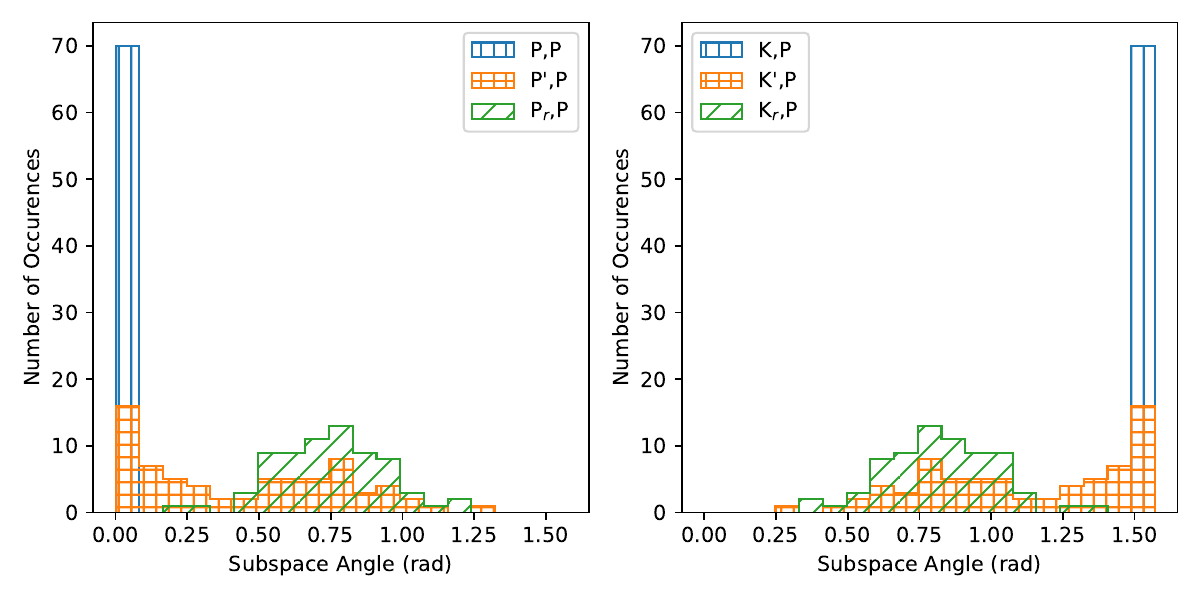}
	\caption{\label{fig:subspace_angles} The subspace angles between various projection matrices (onto dominant modes on the right, onto a robust subspace on the left) and $\mathbf{P}$, the projection onto the dominant error modes determined from the full quadratic model. The number of principle angles with value 0 is the dimension of overlap between the subspaces. Angles with value $\pi/2$ indicate overlap with the subspace orthogonal to $\mathbf{P}$. $\mathbf{P'}$ (the space of dominant modes derived from the approximate model) overlaps with $\mathbf{P}$ (the space of dominant modes derived from the full model) significantly more than random. Crucially, $\mathbf{K'}$ (the robust subspace from the approximate model) overlaps with the subspace orthogonal to $\mathbf{P}$'s significantly more than random, which is why data projected onto this subspace is still robust to wavefront error.}
\end{center}
\end{figure*}

To understand why this is the case, we analyze the subspaces spanned by the identified optimal projection matrices. We define $\mathbf{P}$ as the projection onto the dominant modes of $\mathbf{A_q}$, $\mathbf{P'}$ as the projection onto the dominant modes of $\mathbf{A_q'}$, and $\mathbf{P_r}$ as a random projection matrix the same shape as $\mathbf{P'}$. We also define $\mathbf{K}$ as the projection onto the remaining modes (the robust subspace) of the full model, $\mathbf{K'}$ as the projection onto the robust subspace of the approximate model, and $\mathbf{K_r}$ as a random projection matrix the same shape as $\mathbf{K'}$. We then calculate the subspace angles \citep{jordan_subspace_angles} between each of these projection matrices and $\mathbf{P}$ using the function scipy.linalg.subspace\_angles. These subspace angles provide an indication of how much the subspace spanned by each of these projection matrices overlaps with the subspace spanned by the dominant modes identified by the full model. The results are shown in Figure \ref{fig:subspace_angles}.

The number of principle angles with value 0 is the dimension of overlap between the subspaces. As expected, the subspace angles between $\mathbf{P}$ and itself are all 0, meaning it overlaps completely with itself. Also as expected, the angles between $\mathbf{K}$ and $\mathbf{P}$ are all $\pi/2$, as $\mathbf{K}$ is orthogonal to $\mathbf{P}$. Both of the random matrices have a random distribution of angles with $\mathbf{P}$ centered around $\pi/4$. Meanwhile, $\mathbf{P'}$ (the space of dominant modes derived from the approximate model) overlaps with $\mathbf{P}$ (the space of dominant modes derived from the full model) significantly more than random. Crucially, $\mathbf{K'}$ (the robust subspace from the approximate model) overlaps with $\mathbf{P}$ significantly less than random, and with the subspace orthogonal to $\mathbf{P}$'s significantly more than random, which is why data projected onto this subspace is still robust to wavefront error. This result shows why the approximate model, despite poorly predicting the detector intensity response, is nevertheless useful for identifying a subspace that overlaps significantly with the robust subspace of the full model.

\begin{figure*}[!ht]
\begin{center}
	\includegraphics[scale = 0.75]{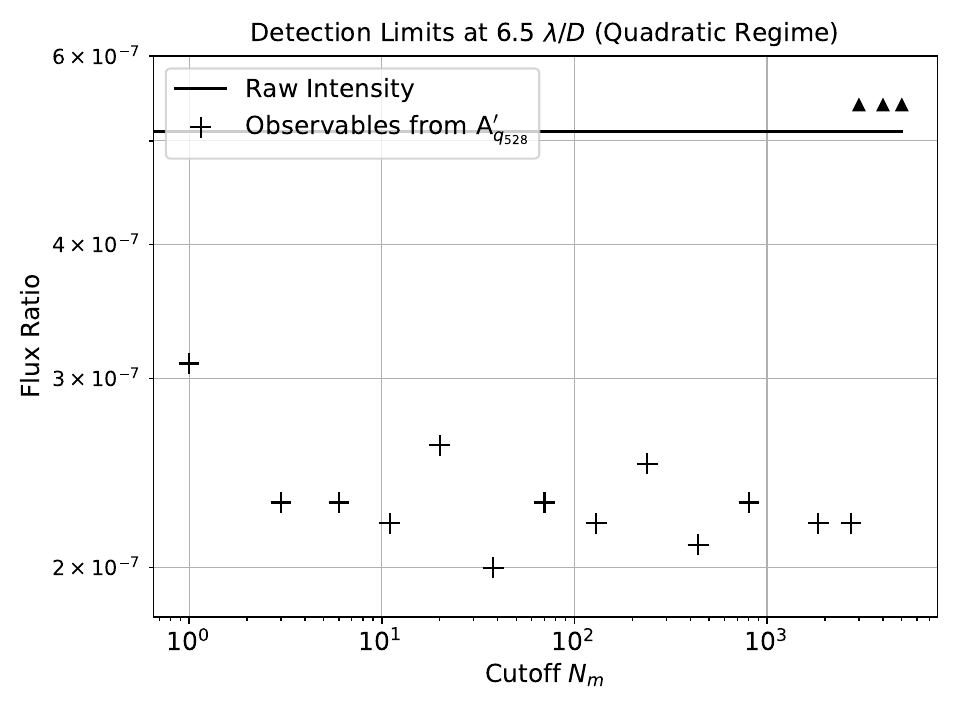}
	\caption{\label{fig:fr_limits_528} Quadratic regime flux ratio detection limits ($\mathrm{FPR}=0.01$, $\mathrm{TPR}=0.90$), to two significant figures, as a function of cutoff mode, for a companion at $6.5\,\ \lambda/D$. Upward triangles indicate that a projection matrix with the specified cutoff mode performs  worse than using the raw intensity, which occurs when the modes the majority of the planet signal overlaps with have also been projected out. The optimal cutoff mode is 38, which results in a detection limit of $2.0 \times 10^{-7}$. Unshowable in log-log scale is the detection limit with $N_m=0$, which, with observables, is $4.9 \times 10^{-7}$. This is, as expected from the fact that no error modes are removed, close to the raw intensity detection limit of $5.1 \times 10^{-7}$.}
\end{center}
\end{figure*}

We can thus use this approximation with all 528 Zernikes in our model to analyze spatial separations beyond the $\sim 5 \,\ \lambda/D$ spanned by the first 100 Zernikes. To demonstrate this, we build $\mathbf{A_{q_{528}}'}$ according to Equation \ref{eq:A_q_prime}, and perform detection tests at a separation of $6.5\,\ \lambda/D$. The results are shown in Figure \ref{fig:fr_limits_528}.

Our tests show that the approximation $\mathbf{A_{q_{528}}'}$ can successfully increase signal-to-noise at spatial separations beyond the original regime of validity of $\mathbf{A_q}$. Thus, even though the input dimension of the quadratic model scales cumbersomely with the number of basis vectors, an approximation considering only norm-squared terms can still be used to find observables that are robust to quadratic wavefront error, and thus provide detection gains at farther spatial separations of interest.


\bibliography{biblio}{}

\begin{thebibliography}{}
\expandafter\ifx\csname natexlab\endcsname\relax\def\natexlab#1{#1}\fi
\providecommand{\url}[1]{\href{#1}{#1}}
\providecommand{\dodoi}[1]{doi:~\href{http://doi.org/#1}{\nolinkurl{#1}}}
\providecommand{\doeprint}[1]{\href{http://ascl.net/#1}{\nolinkurl{http://ascl.net/#1}}}
\providecommand{\doarXiv}[1]{\href{https://arxiv.org/abs/#1}{\nolinkurl{https://arxiv.org/abs/#1}}}

\bibitem[{{Astropy Collaboration} {et~al.}(2013){Astropy Collaboration}, {Robitaille}, {Tollerud}, {Greenfield}, {Droettboom}, {Bray}, {Aldcroft}, {Davis}, {Ginsburg}, {Price-Whelan}, {Kerzendorf}, {Conley}, {Crighton}, {Barbary}, {Muna}, {Ferguson}, {Grollier}, {Parikh}, {Nair}, {Unther}, {Deil}, {Woillez}, {Conseil}, {Kramer}, {Turner}, {Singer}, {Fox}, {Weaver}, {Zabalza}, {Edwards}, {Azalee Bostroem}, {Burke}, {Casey}, {Crawford}, {Dencheva}, {Ely}, {Jenness}, {Labrie}, {Lim}, {Pierfederici}, {Pontzen}, {Ptak}, {Refsdal}, {Servillat}, \& {Streicher}}]{astropy:2013}
{Astropy Collaboration}, {Robitaille}, T.~P., {Tollerud}, E.~J., {et~al.} 2013, Astronomy and Astrophysics, 558, A33, \dodoi{10.1051/0004-6361/201322068}

\bibitem[{{Baudoz} {et~al.}(2006){Baudoz}, {Boccaletti}, {Baudrand}, \& {Rouan}}]{Baudoz_2006}
{Baudoz}, P., {Boccaletti}, A., {Baudrand}, J., \& {Rouan}, D. 2006, in IAU Colloq. 200: Direct Imaging of Exoplanets: Science \& Techniques, ed. C.~{Aime} \& F.~{Vakili}, 553--558, \dodoi{10.1017/S174392130600994X}

\bibitem[{Bloemhof(2002)}]{Bloemhof_2002}
Bloemhof, E.~E. 2002, The Astrophysical Journal, 582, L59, \dodoi{10.1086/346100}

\bibitem[{{Cantalloube} {et~al.}(2022){Cantalloube}, {Christiaens}, {Cantero}, {Nasedkin}, {Cioppa}, {Absil}, {Bonse}, {Delorme}, {Gomez-Gonzalez}, {Juillard}, {Mazoyer}, {Samland}, {Ruffio}, \& {Van Droogenbroeck}}]{cantalloube_2022}
{Cantalloube}, F., {Christiaens}, V., {Cantero}, C., {et~al.} 2022, in Society of Photo-Optical Instrumentation Engineers (SPIE) Conference Series, Vol. 12185, Adaptive Optics Systems VIII, ed. L.~{Schreiber}, D.~{Schmidt}, \& E.~{Vernet}, 1218505, \dodoi{10.1117/12.2627968}

\bibitem[{{Ceau} {et~al.}(2019){Ceau}, {Mary}, {Greenbaum}, {Martinache}, {Sivaramakrishnan}, {Laugier}, \& {N'Diaye}}]{ceau}
{Ceau}, A., {Mary}, D., {Greenbaum}, A., {et~al.} 2019, arXiv e-prints, arXiv:1908.03130.
\newblock \doarXiv{1908.03130}

\bibitem[{Flasseur {et~al.}(2018)Flasseur, Denis, Thiébaut, \& Langlois}]{flasseur}
Flasseur, O., Denis, L., Thiébaut, E., \& Langlois, M. 2018, Astronomy and Astrophysics, 618, \dodoi{10.1051/0004-6361/201832745}

\bibitem[{Groff {et~al.}(2015)Groff, Riggs, Kern, \& Kasdin}]{groff_2015}
Groff, T.~D., Riggs, A. J.~E., Kern, B., \& Kasdin, N.~J. 2015, Journal of Astronomical Telescopes, Instruments, and Systems, 2, 1 , \dodoi{10.1117/1.JATIS.2.1.011009}

\bibitem[{Guyon {et~al.}(2006)Guyon, Pluzhnik, Kuchner, Collins, \& Ridgway}]{Guyon_2006}
Guyon, O., Pluzhnik, E.~A., Kuchner, M.~J., Collins, B., \& Ridgway, S.~T. 2006, The Astrophysical Journal Supplement Series, 167, 81–99, \dodoi{10.1086/507630}

\bibitem[{{Haffert} {et~al.}(2023){Haffert}, {Males}, {Ahn}, {Van Gorkom}, {Guyon}, {Close}, {Long}, {Hedglen}, {Schatz}, {Kautz}, {Lumbres}, {Rodack}, {Knight}, \& {Miller}}]{haffert_2023}
{Haffert}, S.~Y., {Males}, J.~R., {Ahn}, K., {et~al.} 2023, \aap, 673, A28, \dodoi{10.1051/0004-6361/202244960}

\bibitem[{Harris {et~al.}(2020)Harris, Millman, van~der Walt, Gommers, Virtanen, Cournapeau, Wieser, Taylor, Berg, Smith, Kern, Picus, Hoyer, van Kerkwijk, Brett, Haldane, del R{\'{i}}o, Wiebe, Peterson, G{\'{e}}rard-Marchant, Sheppard, Reddy, Weckesser, Abbasi, Gohlke, \& Oliphant}]{harris2020array}
Harris, C.~R., Millman, K.~J., van~der Walt, S.~J., {et~al.} 2020, Nature, 585, 357, \dodoi{10.1038/s41586-020-2649-2}

\bibitem[{Hunter(2007)}]{Hunter:2007_matplotlib}
Hunter, J.~D. 2007, Computing in Science \& Engineering, 9, 90, \dodoi{10.1109/MCSE.2007.55}

\bibitem[{{Ireland}(2013)}]{ireland_2013}
{Ireland}, M.~J. 2013, \mnras, 433, 1718, \dodoi{10.1093/mnras/stt859}

\bibitem[{Jensen-Clem {et~al.}(2017)Jensen-Clem, Mawet, Gonzalez, Absil, Belikov, Currie, Kenworthy, Marois, Mazoyer, Ruane, Tanner, \& Cantalloube}]{Jensen_Clem_2017}
Jensen-Clem, R., Mawet, D., Gonzalez, C. A.~G., {et~al.} 2017, The Astronomical Journal, 155, 19, \dodoi{10.3847/1538-3881/aa97e4}

\bibitem[{Jordan(1875)}]{jordan_subspace_angles}
Jordan, C. 1875, Bulletin de la Soci\'et\'e Math\'ematique de France, 3, 103, \dodoi{10.24033/bsmf.90}

\bibitem[{Kasdin \& Braems(2006)}]{Kasdin_2006}
Kasdin, N.~J., \& Braems, I. 2006, The Astrophysical Journal, 646, 1260, \dodoi{10.1086/505017}

\bibitem[{Kasdin {et~al.}(2020)Kasdin, Bailey, Mennesson, Zellem, Ygouf, Rhodes, Luchik, Zhao, Riggs, Seo, Krist, Kern, Tang, Nemati, Groff, Zimmerman, Macintosh, Turnbull, Debes, Douglas, \& Lupu}]{kasdin_2020}
Kasdin, N.~J., Bailey, V.~P., Mennesson, B., {et~al.} 2020, in Space Telescopes and Instrumentation 2020: Optical, Infrared, and Millimeter Wave, ed. M.~Lystrup, M.~D. Perrin, N.~Batalha, N.~Siegler, \& E.~C. Tong, Vol. 11443, International Society for Optics and Photonics (SPIE), 300 -- 313.
\newblock \url{https://doi.org/10.1117/12.2562997}

\bibitem[{{Krist}(2020)}]{krist_os9}
{Krist}, J. 2020, {WFIRST CGI OS9 Time Series Simulations (Hybrid Lyot Coronagraph, Band 1)}, \url{https://wfirst.ipac.caltech.edu/sims/Coronagraph_public_images.html#CGI_OS9}

\bibitem[{Krist {et~al.}(2023)Krist, Steeves, Dube, Riggs, Kern, Marx, Cady, Zhou, Poberezhskiy, Baker, McGuire, Nemati, Kuan, Mennesson, Trauger, Saini, \& Rafels}]{krist_2023}
Krist, J.~E., Steeves, J.~B., Dube, B.~D., {et~al.} 2023, Journal of Astronomical Telescopes, Instruments, and Systems, 9, 045002, \dodoi{10.1117/1.JATIS.9.4.045002}

\bibitem[{{Lafreni{\`e}re} {et~al.}(2007){Lafreni{\`e}re}, {Marois}, {Doyon}, {Nadeau}, \& {Artigau}}]{lafreniere_loci}
{Lafreni{\`e}re}, D., {Marois}, C., {Doyon}, R., {Nadeau}, D., \& {Artigau}, {\'E}. 2007, Astrophysical Journal, 660, 770, \dodoi{10.1086/513180}

\bibitem[{Laginja {et~al.}(2019)Laginja, Leboulleux, Pueyo, Soummer, Sauvage, Mugnier, Coyle, Knight, Laurent, Por, \& Noss}]{laginja_2019}
Laginja, I., Leboulleux, L., Pueyo, L., {et~al.} 2019, in Techniques and Instrumentation for Detection of Exoplanets IX, Vol. 11117, International Society for Optics and Photonics (SPIE), 382 -- 396, \dodoi{10.1117/12.2530300}

\bibitem[{Leboulleux {et~al.}(2018)Leboulleux, Sauvage, Pueyo, Fusco, Soummer, Mazoyer, Sivaramakrishnan, N'Diaye, \& Fauvarque}]{leboulleux2018pairbased}
Leboulleux, L., Sauvage, J.-F., Pueyo, L., {et~al.} 2018, Pair-based Analytical model for Segmented Telescopes Imaging from Space ({PASTIS}) for sensitivity analysis.
\newblock \doarXiv{1807.00870}

\bibitem[{Marois {et~al.}(2006)Marois, Lafreniere, Doyon, Macintosh, \& Nadeau}]{Marois_2006}
Marois, C., Lafreniere, D., Doyon, R., Macintosh, B., \& Nadeau, D. 2006, The Astrophysical Journal, 641, 556, \dodoi{10.1086/500401}

\bibitem[{{Martinache}(2010)}]{martinache2010}
{Martinache}, F. 2010, \apj, 724, 464, \dodoi{10.1088/0004-637X/724/1/464}

\bibitem[{Nemati {et~al.}(2020)Nemati, Stahl, Stahl, Ruane, \& Sheldon}]{nemati_2020}
Nemati, B., Stahl, H.~P., Stahl, M.~T., Ruane, G. J.~J., \& Sheldon, L.~J. 2020, Journal of Astronomical Telescopes, Instruments, and Systems, 6, 1 , \dodoi{10.1117/1.JATIS.6.3.039002}

\bibitem[{Noll(1976)}]{Noll}
Noll, R.~J. 1976, J. Opt. Soc. Am., 66, 207, \dodoi{10.1364/JOSA.66.000207}

\bibitem[{{Perrin} {et~al.}(2003){Perrin}, {Sivaramakrishnan}, {Makidon}, {Oppenheimer}, \& {Graham}}]{perrin_2003}
{Perrin}, M.~D., {Sivaramakrishnan}, A., {Makidon}, R.~B., {Oppenheimer}, B.~R., \& {Graham}, J.~R. 2003, \apj, 596, 702, \dodoi{10.1086/377689}

\bibitem[{Pogorelyuk {et~al.}(2019)Pogorelyuk, Kasdin, \& Rowley}]{Pogorelyuk_2019}
Pogorelyuk, L., Kasdin, N.~J., \& Rowley, C.~W. 2019, The Astrophysical Journal, 881, 126, \dodoi{10.3847/1538-4357/ab2ecf}

\bibitem[{{Pope} {et~al.}(2021){Pope}, {Pueyo}, {Xin}, \& {Tuthill}}]{pope2021}
{Pope}, B. J.~S., {Pueyo}, L., {Xin}, Y., \& {Tuthill}, P.~G. 2021, \apj, 907, 40, \dodoi{10.3847/1538-4357/abcb00}

\bibitem[{{Price-Whelan} {et~al.}(2018){Price-Whelan}, {Sip{\H{o}}cz}, {G{\"u}nther}, {Lim}, {Crawford}, {Conseil}, {Shupe}, {Craig}, {Dencheva}, {Ginsburg}, {VanderPlas}, {Bradley}, {P{\'e}rez-Su{\'a}rez}, {de Val-Borro}, {Paper Contributors}, {Aldcroft}, {Cruz}, {Robitaille}, {Tollerud}, {Coordination Committee}, {Ardelean}, {Babej}, {Bach}, {Bachetti}, {Bakanov}, {Bamford}, {Barentsen}, {Barmby}, {Baumbach}, {Berry}, {Biscani}, {Boquien}, {Bostroem}, {Bouma}, {Brammer}, {Bray}, {Breytenbach}, {Buddelmeijer}, {Burke}, {Calderone}, {Cano Rodr{\'\i}guez}, {Cara}, {Cardoso}, {Cheedella}, {Copin}, {Corrales}, {Crichton}, {D{\textquoteright}Avella}, {Deil}, {Depagne}, {Dietrich}, {Donath}, {Droettboom}, {Earl}, {Erben}, {Fabbro}, {Ferreira}, {Finethy}, {Fox}, {Garrison}, {Gibbons}, {Goldstein}, {Gommers}, {Greco}, {Greenfield}, {Groener}, {Grollier}, {Hagen}, {Hirst}, {Homeier}, {Horton}, {Hosseinzadeh}, {Hu}, {Hunkeler}, {Ivezi{\'c}}, {Jain}, {Jenness}, {Kanarek}, {Kendrew}, {Kern}, {Kerzendorf}, {Khvalko},
  {King}, {Kirkby}, {Kulkarni}, {Kumar}, {Lee}, {Lenz}, {Littlefair}, {Ma}, {Macleod}, {Mastropietro}, {McCully}, {Montagnac}, {Morris}, {Mueller}, {Mumford}, {Muna}, {Murphy}, {Nelson}, {Nguyen}, {Ninan}, {N{\"o}the}, {Ogaz}, {Oh}, {Parejko}, {Parley}, {Pascual}, {Patil}, {Patil}, {Plunkett}, {Prochaska}, {Rastogi}, {Reddy Janga}, {Sabater}, {Sakurikar}, {Seifert}, {Sherbert}, {Sherwood-Taylor}, {Shih}, {Sick}, {Silbiger}, {Singanamalla}, {Singer}, {Sladen}, {Sooley}, {Sornarajah}, {Streicher}, {Teuben}, {Thomas}, {Tremblay}, {Turner}, {Terr{\'o}n}, {van Kerkwijk}, {de la Vega}, {Watkins}, {Weaver}, {Whitmore}, {Woillez}, {Zabalza}, \& {Contributors}}]{astropy:2018}
{Price-Whelan}, A.~M., {Sip{\H{o}}cz}, B.~M., {G{\"u}nther}, H.~M., {et~al.} 2018, Astronomical Journal, 156, 123, \dodoi{10.3847/1538-3881/aabc4f}

\bibitem[{{Pueyo}(2016)}]{pueyo_klip}
{Pueyo}, L. 2016, Astrophysical Journal, 824, 117, \dodoi{10.3847/0004-637X/824/2/117}

\bibitem[{Riggs {et~al.}(2018)Riggs, Ruane, Sidick, Coker, Kern, \& Shaklan}]{riggs_falco}
Riggs, A. J.~E., Ruane, G., Sidick, E., {et~al.} 2018, in Space Telescopes and Instrumentation 2018: Optical, Infrared, and Millimeter Wave, ed. M.~Lystrup, H.~A. MacEwen, G.~G. Fazio, N.~Batalha, N.~Siegler, \& E.~C. Tong, Vol. 10698, International Society for Optics and Photonics (SPIE), 878 -- 888, \dodoi{10.1117/12.2313812}

\bibitem[{{Soummer} {et~al.}(2012){Soummer}, {Pueyo}, \& {Larkin}}]{soummer_klip}
{Soummer}, R., {Pueyo}, L., \& {Larkin}, J. 2012, Astrophysical Journal, Letters, 755, L28, \dodoi{10.1088/2041-8205/755/2/L28}

\bibitem[{Sparks \& Ford(2002)}]{Sparks_2002}
Sparks, W.~B., \& Ford, H.~C. 2002, The Astrophysical Journal, 578, 543, \dodoi{10.1086/342401}

\bibitem[{{Traub} \& {Oppenheimer}(2010)}]{traub_oppenheimer}
{Traub}, W.~A., \& {Oppenheimer}, B.~R. 2010, {Direct Imaging of Exoplanets}, ed. S.~{Seager}, 111--156

\bibitem[{Tyson(2000)}]{tyson}
Tyson, R.~K. 2000, Introduction to Adaptive Optics (Bellingham: The Society of Photo-optical Instrumetation Engineers)

\bibitem[{{Virtanen} {et~al.}(2020){Virtanen}, {Gommers}, {Oliphant}, {Haberland}, {Reddy}, {Cournapeau}, {Burovski}, {Peterson}, {Weckesser}, {Bright}, {van der Walt}, {Brett}, {Wilson}, {Jarrod Millman}, {Mayorov}, {Nelson}, {Jones}, {Kern}, {Larson}, {Carey}, {Polat}, {Feng}, {Moore}, {Vand erPlas}, {Laxalde}, {Perktold}, {Cimrman}, {Henriksen}, {Quintero}, {Harris}, {Archibald}, {Ribeiro}, {Pedregosa}, {van Mulbregt}, \& {Contributors}}]{2020SciPy-NMeth}
{Virtanen}, P., {Gommers}, R., {Oliphant}, T.~E., {et~al.} 2020, Nature Methods, 17, 261, \dodoi{https://doi.org/10.1038/s41592-019-0686-2}

\bibitem[{Vogt {et~al.}(2011)Vogt, Martinache, Guyon, Yoshikawa, Yokochi, Garrel, \& Matsuo}]{Vogt_2011}
Vogt, F. P.~A., Martinache, F., Guyon, O., {et~al.} 2011, Publications of the Astronomical Society of the Pacific, 123, 1434–1441, \dodoi{10.1086/663723}

\bibitem[{{Ygouf} {et~al.}(2016){Ygouf}, {Zimmerman}, {Pueyo}, {Soummer}, {Perrin}, {Mennesson}, {Krist}, {Vasisht}, {Nemati}, \& {Macintosh}}]{ygouf_2016}
{Ygouf}, M., {Zimmerman}, N.~T., {Pueyo}, L., {et~al.} 2016, in Society of Photo-Optical Instrumentation Engineers (SPIE) Conference Series, Vol. 9904, Space Telescopes and Instrumentation 2016: Optical, Infrared, and Millimeter Wave, ed. H.~A. {MacEwen}, G.~G. {Fazio}, M.~{Lystrup}, N.~{Batalha}, N.~{Siegler}, \& E.~C. {Tong}, 99045M, \dodoi{10.1117/12.2231581}

\end{thebibliography}
\bibliographystyle{aasjournal}



\end{document}